\documentclass[11pt]{article}
\pdfoutput=1
\usepackage{color}
\usepackage{cite}
\usepackage{graphicx}
\usepackage{multirow}
\usepackage{float}

\usepackage{amssymb,amsmath}

\def\ba{\begin{array}{c}}
\def\bat{\begin{array}{cc}}
\def\bat{\begin{array}{cc}}
\def\ea{\end{array}}
\def\bat{\begin{array}{cc}}
\def\batt{\begin{array}{ccc}}
\newcommand{\bel}[1]{\be\label{#1}}

\newcommand{\be}{\begin{equation}}
\newcommand{\ee}{\end{equation}}
\def\beqn{\begin{eqnarray}}
\def\eeqn{\end{eqnarray}}

\definecolor{nicered}{rgb}{1.0,0.0,0.2}

\usepackage{color}

\begin{document}

\title{
\begin{flushright}\vbox{\normalsize \mbox{}\vskip -6cm
FTUV/13-0913 \\[-3pt] IFIC/13-63 \\[-3pt] LA-UR-13-27104
 \\[-3pt]
}
\end{flushright}\vskip 45pt
{\bf  Lepton flavor violation in the Higgs sector and the role of hadronic $\tau$-lepton decays }}
\bigskip

\author{Alejandro Celis$^{1}$, Vincenzo Cirigliano$^{2}$ and Emilie Passemar$^{2}$\\[15pt]
{$^1$\small IFIC, Universitat de Val\`encia -- CSIC, Apt. Correus 22085, E-46071 Val\`encia, Spain}\\
{$^2$\small Theoretical Division, Los Alamos National Laboratory, Los Alamos,  NM 87545, USA}}

\date{}
\maketitle
\bigskip \bigskip

\begin{abstract}
\noindent
It has been pointed out recently that current low-energy constraints still allow for sizable flavor-changing decay rates of the $125$~GeV boson into leptons, $h \rightarrow \tau \ell$ ($\ell= e, \mu$).  In this work we discuss the role of hadronic $\tau$-lepton decays in probing lepton flavor violating couplings in the Higgs sector.  
At low energy, the effective Higgs coupling to gluons induced by heavy quarks contributes to
hadronic $\tau$-decays, establishing a direct connection with the relevant process at the LHC, $pp (gg) \rightarrow h \rightarrow \tau \ell$.     Semileptonic transitions like $\tau \rightarrow \ell \pi \pi$ are sensitive to flavor-changing scalar couplings while decays such as $\tau \rightarrow \ell \eta^{(\prime)}$ probe pseudoscalar couplings, thus providing a useful low-energy handle to disentangle  possible Higgs flavor violating signals at the LHC.   As part of our analysis,  we provide 
an appropriate description of all the relevant hadronic matrix elements needed to describe Higgs mediated $\tau \rightarrow \ell \pi \pi$ transitions, improving over previous treatments in the literature.     
\end{abstract}

\newpage

\section{Introduction}
 With the discovery of a new boson with mass close to $125$~GeV, here referred as $h(125)$, a new era in the understanding of the electroweak symmetry breaking (EWSB) mechanism has started. Current experimental data already indicate that this boson is related to the origin of particle masses and its properties are so far in good agreement with those of the standard model (SM) Higgs boson~\cite{CMS:yva,ATLAS:2013sla}.  The spin-parity of the new particle are consistent with the assignment $J^{P}=0^{+}$, other possibilities being strongly disfavoured.    Global fits of the ATLAS, CMS and Tevatron data also find that the couplings of this boson to the gauge vector bosons ($\gamma, g, W^{\pm}, Z$) and the third family of fermions ($t, b, \tau$) are compatible with the SM expectation~\cite{Cheung:2013kla,Ellis:2013lra,Falkowski:2013dza,Giardino:2013bma}.
 
 Searches for lepton flavor violating (LFV) Higgs decays at the LHC offer an interesting possibility to test for new physics effects that could have escaped current experimental low-energy constraints~\cite{Blankenburg:2012ex}.  LFV  effects associated with the scalar sector have been studied considerably in the 
 past~\cite{DiazCruz:1999xe,Black:2002wh,Babu:2002et,
  Brignole:2004ah,Kanemura:2005hr,Paradisi:2005tk,Arganda:2008jj,Herrero:2009tm, Davidson:2010xv, Goudelis:2011un}.    
The recent discovery of the $h(125)$ boson at the LHC has naturally caused renewed interest in this possibility~\cite{Blankenburg:2012ex,Harnik:2012pb,Davidson:2012ds,Crivellin:2013wna,Dery:2013rta,Arhrib:2012ax,Arana-Catania:2013xma}.  In this work we address several questions related to LFV in the Higgs sector:
\begin{itemize}
\item How robust a connection can be made between the LFV Higgs decays and LFV  $\tau$ decays?
\item What is the role of hadronic $\tau$ decays ($\tau \rightarrow \ell \pi \pi, \ell \eta^{(\prime)}, \ldots$) compared to other $\tau$ decays ($\tau \rightarrow \ell \gamma, \ldots$) in probing LFV couplings of the Higgs sector? 
\item What can be said about LFV phenomena within the general two-Higgs-doublet model based on our current knowledge of the $h(125)$ properties?
\end{itemize}
Along the way we provide an appropriate treatment of the form factors 
needed to study hadronic  LFV $\tau$ decays.
These  will be useful for any analysis  of LFV $\tau$ decays,   beyond the specific framework adopted here.   In this work we will not attempt to perform a study of all the available LFV hadronic decay modes. Indeed, just for semileptonic transitions the experimental collaborations have considered at the moment a great variety of hadronic final states $\tau \rightarrow \ell (\pi \pi, \pi K, KK, \eta \pi, \eta \eta)$.  Instead, we focus here on $\tau \rightarrow \ell \pi \pi$ semileptonic transitions for which a better control of the relevant hadronic matrix elements can be achieved. Concerning the $\tau \rightarrow \ell P$ decays, we restrict the discussion to a few modes $P= \pi, \eta, \eta^{\prime}$ for clarity.  The richness of hadronic $\tau$ decay modes could certainly be extremely useful in the future to corroborate any possible LFV signal at the LHC, 
providing complementary information to scrutinize its origin.

\subsection{Motivation}
 Flavor violating couplings of the Higgs boson to leptons arise in many extensions of the SM. 
 If the new physics originates at a scale $\Lambda$ well above the EW scale,  then an effective theory  treatment is justifiable.    Details of the ultraviolet completion of the SM can then be encoded in effective operators containing only the SM degrees of freedom and which can give rise to LFV effects~\cite{Buchmuller:1985jz,Grzadkowski:2010es}.   It is however possible that the new physics (NP) enters at a scale not much higher than the electroweak scale, so that these new degrees of freedom cannot be integrated out.  One possibility is to consider an extended Higgs sector with several scalar fields below the TeV scale and non-diagonal Yukawa couplings in flavor space.   Indeed, it is the case that a simple extension of the SM scalar sector by an additional Higgs doublet, a two-Higgs-doublet model (2HDM), gives rise to flavor-changing neutral currents (FCNCs) at tree-level in the quark and lepton sectors.   Usually a symmetry principle is assumed that forbids such effects~\cite{Gunion:1989we,Branco:2011iw} and allows to evade the stringent bounds coming from the Kaon and $B$-meson precision experiments. While such an approach is well justified or needed for the quark sector, one can be less restrictive in  the lepton sector,  while being consistent with flavor constraints.

The strongest bound on possible LFV Higgs couplings to $\tau-\ell$ ($\ell = e, \mu$) are currently obtained from $\tau \rightarrow \ell \gamma$ decays.   It was first noticed in Ref.~\cite{Blankenburg:2012ex} that present bounds still allow for very large LFV Higgs decay rates $\mathrm{BR}(h \rightarrow \tau \ell) \lesssim 10 \%$.   It was later shown in Refs.~\cite{Harnik:2012pb,Davidson:2012ds} that the LHC prospects in constraining such LFV Higgs couplings are very promising, even with present accumulated data. 
Constraining LFV couplings of the $125$~GeV Higgs directly at the LHC, or finding additional low-energy handles,
becomes even more relevant when one considers the nature of the bound that can be extracted from $\tau \rightarrow \ell \gamma$ decays.  
The effective dipole operator $(\bar \ell  \sigma^{\mu \nu} P_{L,R} \tau ) F_{\mu \nu}$ giving rise to $\tau \rightarrow \ell \gamma$ decays appears at the loop-level and is very sensitive to details of the high energy dynamics.   
Due to the strong chirality suppression of the one-loop diagrams, the dominant contribution to the $\tau \rightarrow \ell \gamma$ decay amplitude arises from two-loop diagrams of the Barr-Zee type~\cite{Chang:1993kw}.  Additional scalars or heavy degrees of freedom belonging to the UV completion of the theory can cause sizable interfering contributions, making it impossible to extract a model independent bound on the scalar LFV couplings.   This  issue is circumvented at the LHC by searching directly for LFV Higgs decays, the Higgs being produced via its coupling to $VV=W^+W^-,ZZ$ in vector-boson fusion and in associated Higgs production with a vector boson~\cite{Harnik:2012pb}, or,  relying on its loop-induced coupling to gluons in the gluon fusion mode~\cite{Davidson:2012ds}.   

\begin{figure}[ht!]
\centering
\includegraphics[width=5cm,height=4.cm]{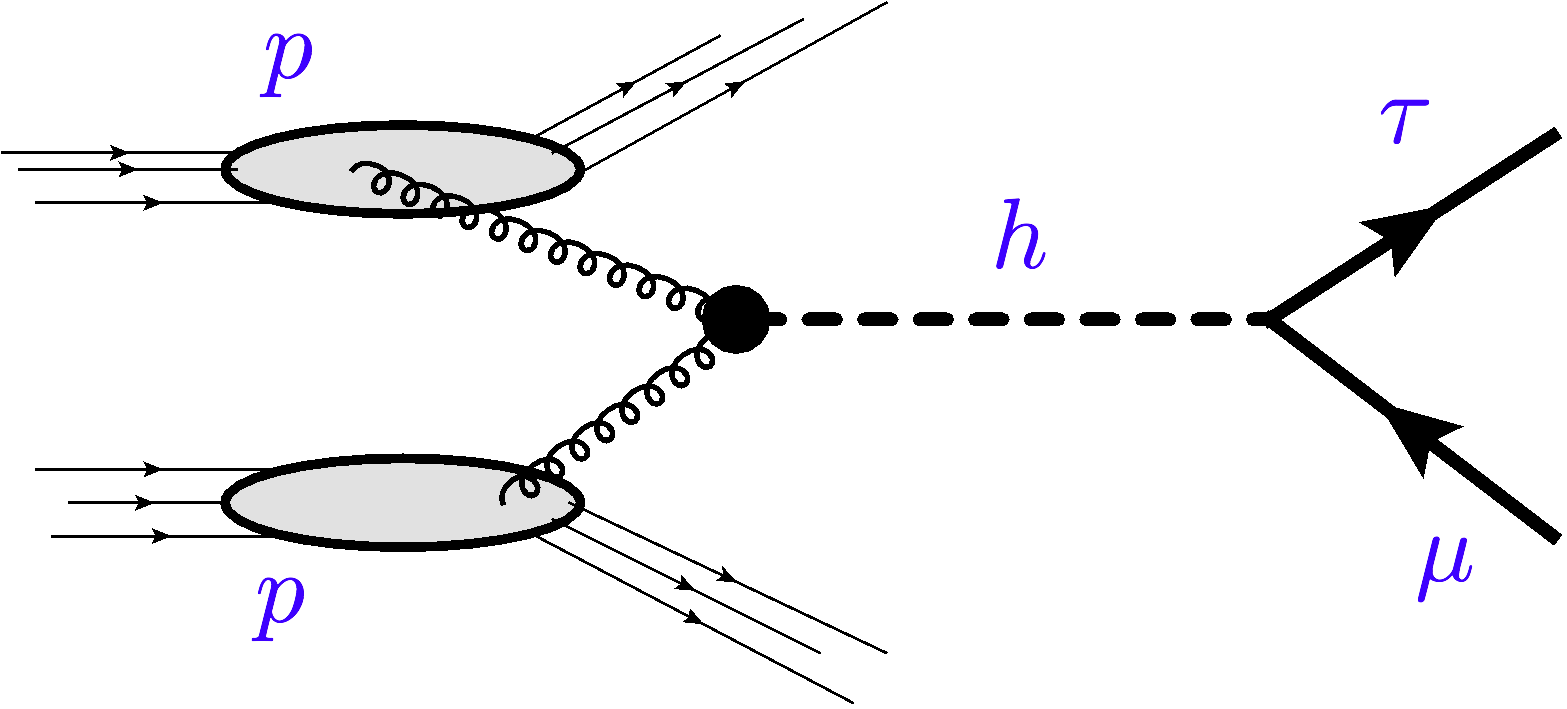}
~~~~~~~~~~~~~~~~~
\includegraphics[width=5cm,height=4cm]{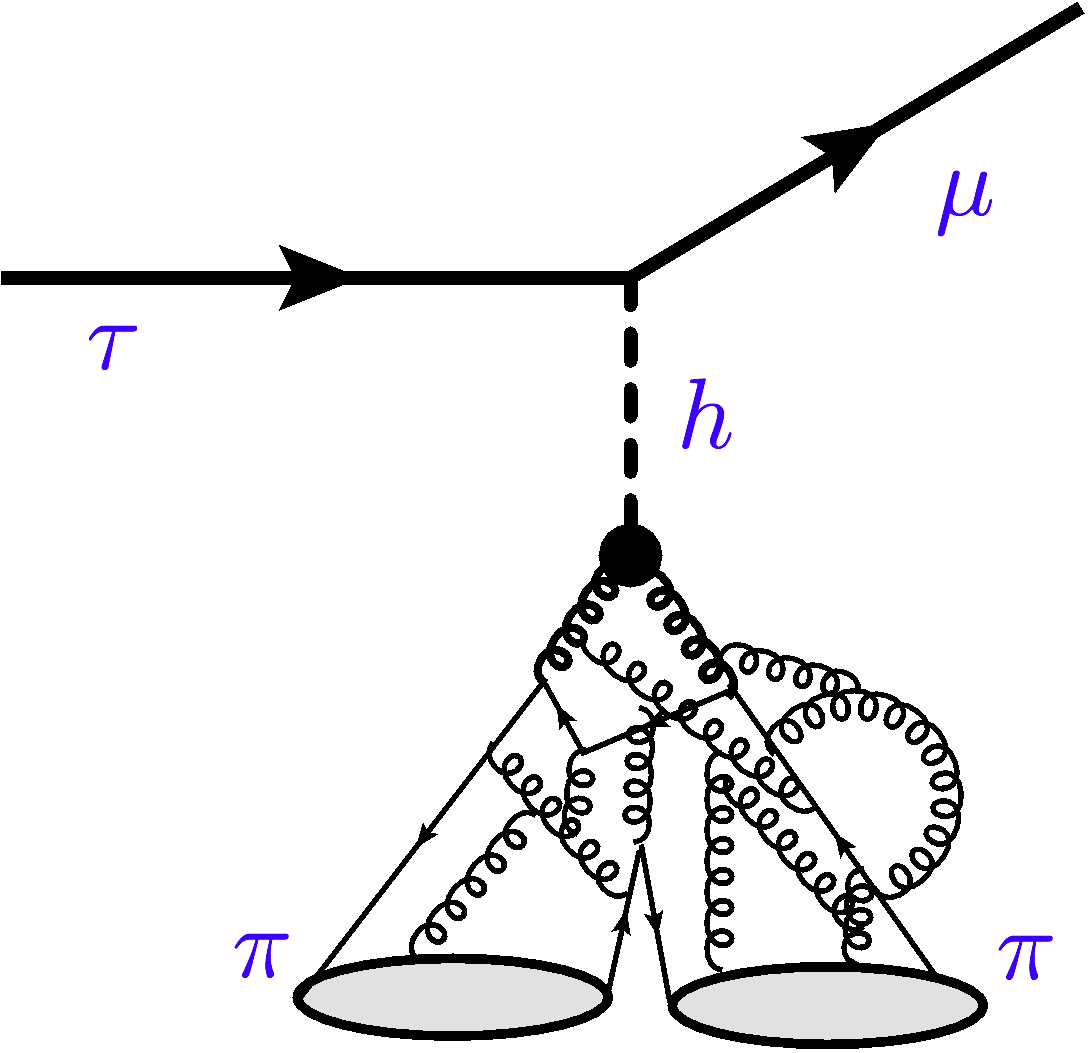}
\caption{\label{taumupipivsLHC} \it \small  Relation between the LHC process $pp (gg) \rightarrow h \rightarrow  \tau \mu$ (left figure) and the semileptonic decay $\tau \rightarrow \mu \pi \pi$ (right figure):  the effective Higgs coupling to gluons enters in both processes.  }
\end{figure}

At low energy, semileptonic $\tau$ decays like $\tau \rightarrow \ell \pi \pi$ ($\pi\pi = \pi^+ \pi^-, \pi^0 \pi^0$) offer a unique opportunity 
to extract a bound on the LFV Higgs couplings which is much less sensitive to details of the high energy dynamics, thus  establishing a model-independent connection with the search for LFV Higgs decays at the LHC.   The same effective coupling of the Higgs to gluons that would give rise to $pp (gg) \rightarrow h \rightarrow \tau \ell$, also enters in the $\tau \rightarrow \ell \pi \pi$ mode though at a much lower energy scale where non-perturbative QCD effects play a major role, see Fig.~\ref{taumupipivsLHC}.  Similarly, the semileptonic decays $\tau \rightarrow \ell P$ (where $P$ is a pseudoscalar meson) establish a connection with the search for LFV decays of a CP-odd Higgs at the LHC.

Calculations of $\tau \rightarrow \ell \pi \pi$ mediated by a Higgs boson with LFV couplings in the literature have mostly considered the scalar-current associated with the Higgs coupling to light quarks, thus neglecting the effective coupling of the Higgs to gluons due to intermediate heavy quarks (with the exception of Ref.~\cite{Petrov:2013vka}, in which a more general EFT analysis including gluon operators is presented).  Moreover, a description of the scalar-current hadronic matrix elements based on leading order predictions of Chiral-Perturbation Theory (ChPT) has been used in these 
works~\cite{Black:2002wh,Kanemura:2005hr,Arganda:2008jj,Petrov:2013vka}.  Such treatment of the hadronic matrix elements is not adequate to deal with $\tau$ decays, for 
which the $\pi \pi $ invariant mass can be as large as $m_\tau - m_\ell$. It was pointed out recently in Ref.~\cite{Daub:2012mu}, within the context of R-parity violating supersymmetry, that by using a more appropriate description of the hadronic matrix elements of the scalar and vector currents, the bounds obtained on the R-parity-violating couplings improve considerably.   

\subsection{Overview of results}
 In this work we provide for the first time a complete description of the $\tau \rightarrow \ell \pi \pi$ mode in the presence of a Higgs boson with LFV couplings.  A detailed discussion of the hadronic matrix elements involved is given.  When relevant we also compare the form factors we obtain with those of previous work.
With these tools in hand, we extract  from $\tau \to \ell \pi \pi$ robust model-independent bounds on LFV couplings of the Higgs.
 The LFV decays $\tau \rightarrow \ell P$ and the relevant hadronic matrix elements in this case are also discussed, leading to bounds on LFV couplings of a CP-odd neutral scalar.

In the context of an extended Higgs sector, we also point out the importance of performing searches for additional Higgs bosons 
in the LFV decay modes $\tau- \mu$ and $\tau -e$ at the LHC. 
Present data constrain the $h(125)$ coupling to vector bosons to be very close to the SM value
 $g_{hVV}  \simeq g_{hVV}^{\mathrm{SM}}$~\cite{CMS:yva,ATLAS:2013sla,Cheung:2013kla}. 
In general two-Higgs-doublet models, any possible LFV  coupling of the $125$~GeV Higgs boson at the end turns out to be suppressed by an accompanying small or vanishing mixing factor $\left(1- (g_{hVV}/g_{hVV}^{\mathrm{SM}})^2 \right)^{1/2}$.  Additional Higgs bosons which would play a minor role in the restoration of perturbative unitarity on the other hand, do not receive this suppression of their LFV couplings.  The search for LFV decays associated to the scalar sector should therefore not be restricted to the $125$ GeV boson.

Our paper is organized as follows:  In Sect.~\ref{sec::intro2} we describe our framework.  
In Sect.~\ref{sec::ffs} we provide a detailed discussion of the hadronic form factors relevant for the description of $\tau \rightarrow \mu \pi \pi$ decays.   In Sect.~\ref{sec:Pheno} we describe the framework used in this work to motivate the discussion of possible LFV effects due to both CP-even and CP-odd Higgs bosons.  We then consider the semileptonic LFV decay $\tau \rightarrow \mu \pi \pi$ ($\pi \pi = \pi^+ \pi^-, \pi^0 \pi^0$) mediated by a CP-even Higgs boson, we discuss the relevance of this process in connection to other LFV transitions accessible at $B$-factories ($\tau \rightarrow \mu \gamma, 3 \mu, \ldots$) as well as for the LHC ($h \rightarrow \tau \mu$).  The phenomenology of a CP-odd Higgs boson is also discussed along the same lines. We give our conclusions in Sect.~\ref{sum}.

\section{Framework}
\label{sec::intro2}

We will consider the following phenomenological Lagrangian that describes the fermionic interactions of a generic extended scalar sector, 
\be \label{laggen}
\mathcal{L} = - m_k \, \bar f_L^{\,k} \,f_R^{\,k}  - \sum_{\varphi}   \, Y^{\varphi}_{ij} \, \left( \bar f_L^{\,i} \, f_{R}^{\,j}  \right) \, \varphi + \mathrm{h.c.}    ~, 
\ee
where $\varphi$ runs over the light  neutral scalars of the theory, the Yukawa couplings can be complex in principle 
and the various physical scalar fields do not need to be CP eigenstates.  
A similar Lagrangian has been considered very recently in Refs.~\cite{Blankenburg:2012ex,Harnik:2012pb} 
to analyze possible flavor violating effects of a CP-even Higgs of mass 125 GeV.  
In the SM there is only one physical CP-even scalar field, $h$, with Yukawa couplings given by $Y^{h}_{ij} = (m_i/v) \delta_{ij}$.   
We will parametrize the deviations from the SM diagonal couplings as $Y^{h}_{ii} = y_i^{h} \, (m_i/v) $ for convenience in the following.   
Since here we are not interested in CP-violating effects we will assume that CP is a good symmetry of the scalar interactions.    
Physical scalars are then CP-eigenstates and the couplings  $Y^{\varphi}_{ij}$ are real for a CP-even Higgs, $\varphi \equiv h$, 
or pure imaginary for a CP-odd Higgs, $\varphi  \equiv A$.     

In Sect.~\ref{sec:Framework} we will discuss how the non-standard Higgs fermion couplings  of  Eq.~\eqref{laggen}
arise within the framework of the  general 2HDM,  or from higher dimensional gauge invariant operators. 
There we will also discuss  in detail the phenomenological  impact of LFV couplings of  the CP-even and CP-odd scalars. 
Here, we outline in general terms the low-energy effects of the non-standard couplings of  Eq. \eqref{laggen} 
and motivate the analysis of hadronic matrix elements to be discussed in Sect.~\ref{sec::ffs}.

\begin{figure}[t!]
\centering
\includegraphics[width=0.28\textwidth]{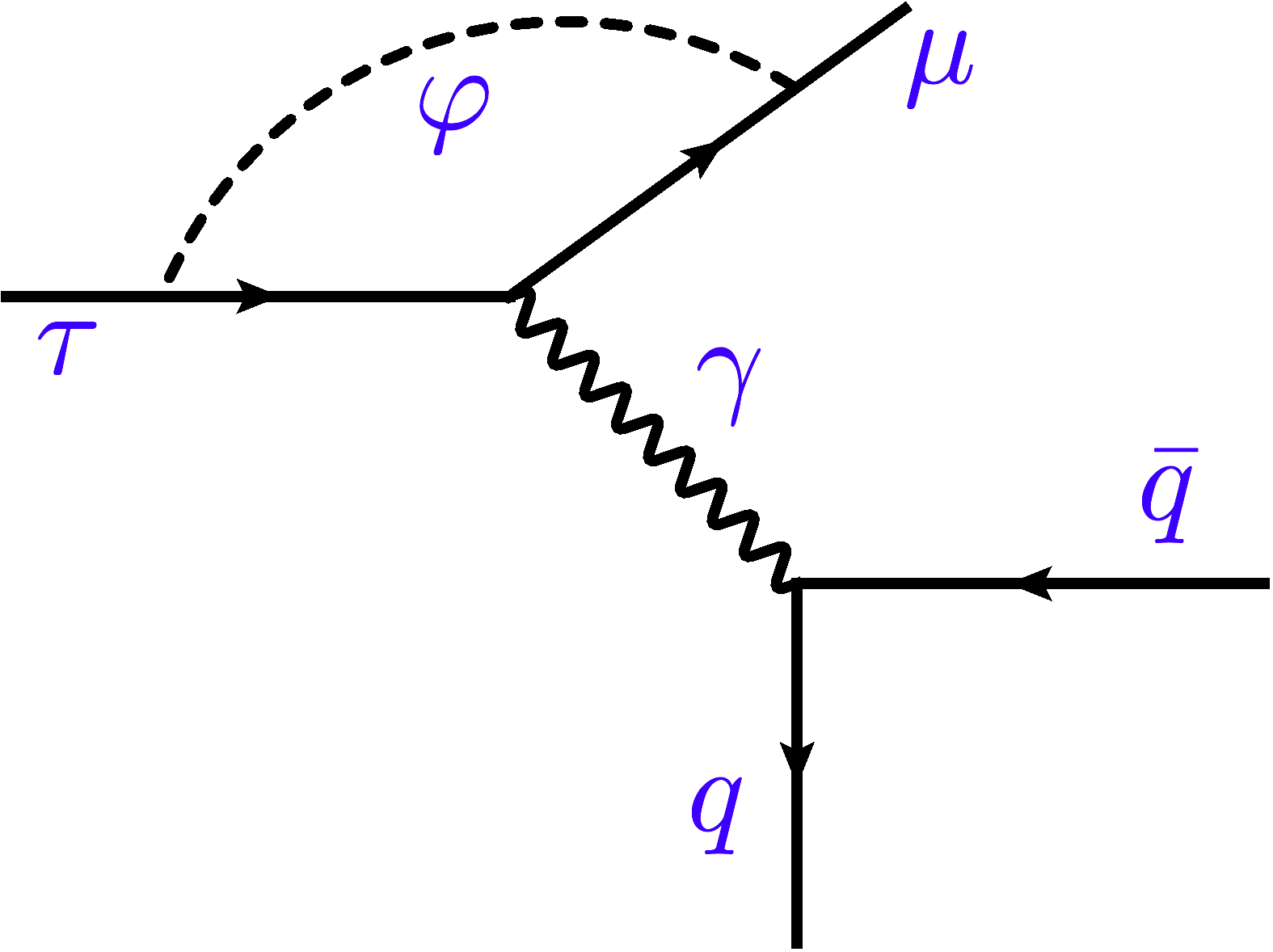}
~~
\includegraphics[width=0.28\textwidth]{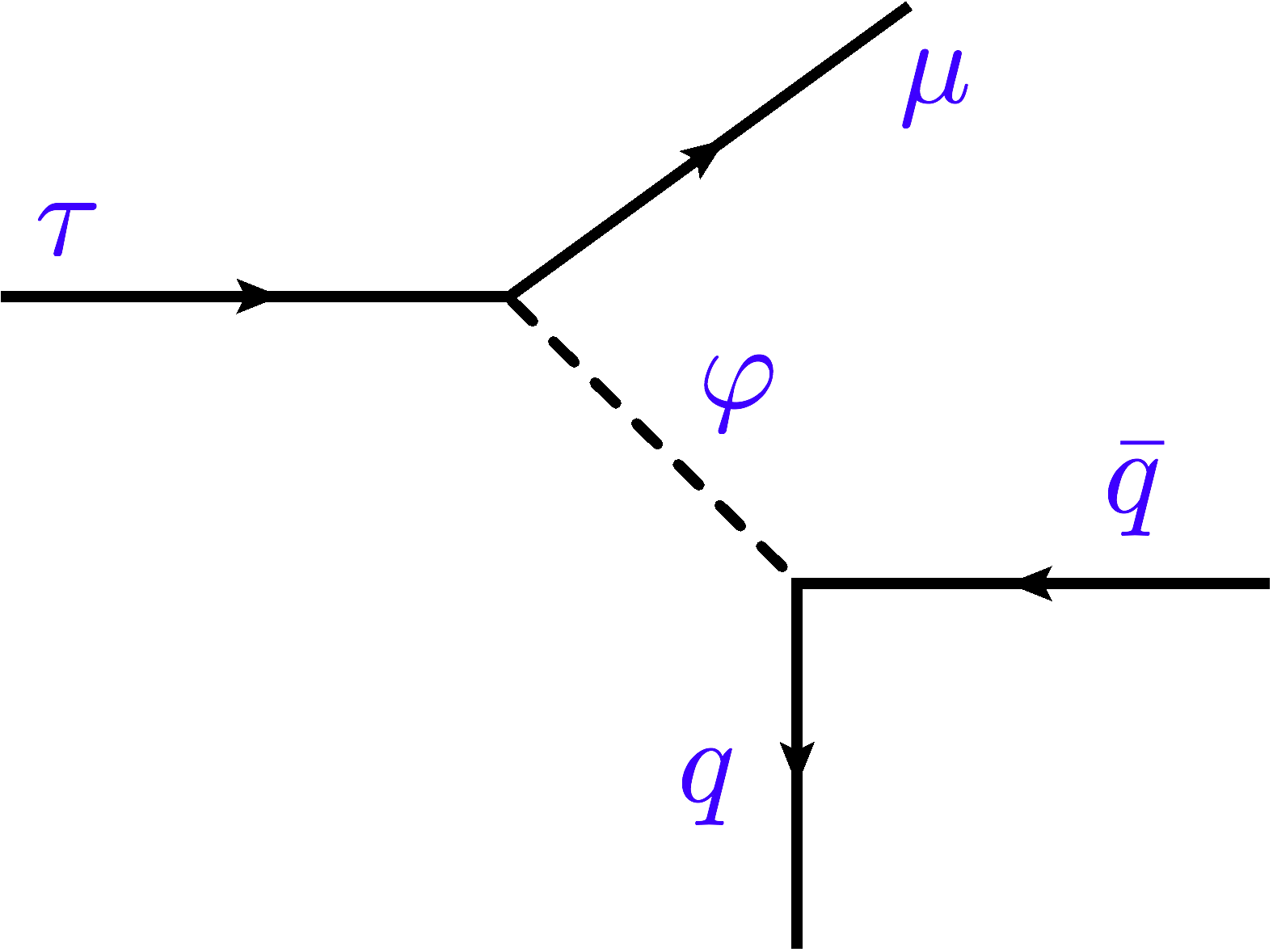}
~~
\includegraphics[width=0.28\textwidth]{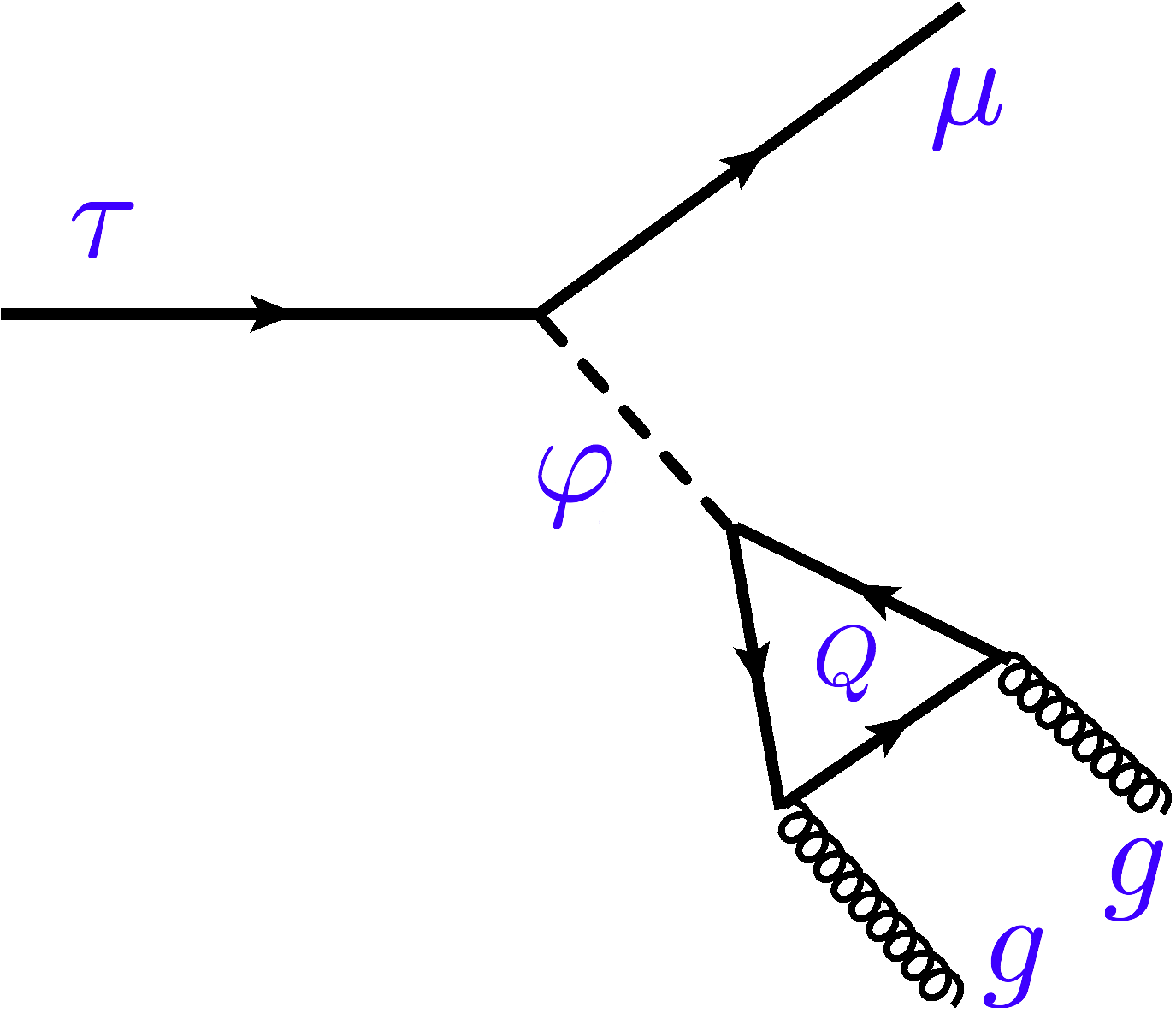}
\caption{\label{fig:integrate} \it \small   Integrating out the Higgs field(s) generates at low-energy several  LFV operator structures: 
dipole (left diagram), scalar four-fermion (center diagram), gluon (right diagram). }
\end{figure}

At low energy, where the Higgs fields can be integrated out,  the  fermion couplings of  Eq. \eqref{laggen} generate a set of 
LFV operators,  as depicted by representative diagrams in Fig.~\ref{fig:integrate}. 
The diagram to the left generates at one-loop the dipole operator 
$(\bar \ell  \sigma^{\mu \nu} P_{L,R} \tau ) F_{\mu \nu}$. Additional two-loop contributions to this operator are not 
shown in Fig.~\ref{fig:integrate} but will be included in the calculation. 
The tree-level diagram in the middle generates a four-fermion operator 
with scalar or pseudoscalar couplings to the light quarks,   $\bar{\ell} (1 \pm \gamma_5)  \tau  \cdot \bar{q} \{ 1, \gamma_5\} q$. 
Finally, the diagram to the right,  through heavy-quarks in the loop generates gluonic operators of the type 
$\bar{\ell} (1 \pm \gamma_5) \tau  \cdot  G G$ and  $\bar{\ell} (1 \pm \gamma_5) \tau  \cdot  G \tilde{G}$.

When considering hadronic LFV decays such as $\tau \to \ell \pi \pi$ or $\tau \to \ell P$ ($P=\pi, \eta, \eta'$) one needs  
the matrix elements of the quark-gluon operators  in the hadronic states. 
In particular, P-even operators will mediate the $\tau \to \ell \pi \pi$  decay and one needs to know 
the relevant two-pion form factors. 
The  dipole operator requires the vector form factor  related to  $\langle \pi \pi |\bar{q} \gamma_\mu q |0 \rangle$ (photon converting in two pions). 
The scalar operator requires the scalar form factors related to   $\langle \pi \pi |\bar{q} q |0 \rangle$. 
The gluon operator requires   $\langle \pi \pi | G G |0 \rangle$, which we will reduce to a  combination of the 
scalar form factors and the  two-pion matrix element of the trace of the  energy-momentum tensor  $\langle \pi \pi | \theta^\mu_\mu  |0 \rangle$
via the trace anomaly relation:
\be
\label{eq:trace}
\theta_{\mu}^{\mu} = - 9 \frac{\alpha_s}{8 \pi} G_{\mu \nu}^{a}  G^{\mu \nu}_{a}  + \sum_{q=u,d,s} m_q \bar q q ~.
\ee
To impose robust bounds on LFV Higgs couplings from $\tau \to \ell \pi \pi$, we need to know the hadronic  matrix elements  with a good  accuracy. 
With this motivation in mind, we now discuss in detail the derivation of  the two-pion matrix elements.

\section{Hadronic form factors for $\tau \rightarrow \ell \pi \pi$ decays}
\label{sec::ffs}
The dipole contribution to the $\tau \to \ell \pi \pi$ decay requires the matrix element
\be
\big\langle \pi^+(p_{\pi^+})\pi^-(p_{\pi^-})\big|\tfrac{1}{2}(\bar
u\gamma^\alpha u-\bar d\gamma^\alpha d)\big|0\big\rangle 
\equiv F_V(s)(p_{\pi^+}-p_{\pi^-})^\alpha,
\label{FVdef}
\ee
with $F_V(s)$ the pion vector form factor. As for the scalar currents and the trace of the  energy-momentum tensor $\theta_\mu^\mu$, the hadronic matrix elements are given by     
\begin{eqnarray}
\big\langle \pi^+(p_{\pi^+})\pi^-(p_{\pi^-})\big | m_u \bar u u+ m_d \bar d d \big|0\big\rangle 
\!\! & \equiv \!\! &\Gamma_\pi (s)~, \nonumber \\
\big\langle \pi^+(p_{\pi^+})\pi^-(p_{\pi^-})\big | m_s \bar s s  \big|0\big\rangle 
\!\! & \equiv \!\! &\Delta_\pi (s)~, \nonumber \\
\big\langle \pi^+(p_{\pi^+})\pi^-(p_{\pi^-})\big | \theta_\mu^\mu  \big|0\big\rangle 
\!\! & \equiv \!\! &\theta_\pi (s)~, 
\label{eq:scalFFdef}
\end{eqnarray}
with $\Gamma_\pi(s)$ and $\Delta_\pi(s)$ the pion scalar form factors and $\theta_\pi(s)$ the form factor related to $\theta_\mu^\mu$. Here
$s$ is the invariant mass squared of the pion pair: $s = \left( p_{\pi^+} + p_{\pi^-} \right)^2 = \left( p_{\tau} - p_{\ell} \right)^2$. 

In what follows, we determine the form factors   by matching a dispersive parameterization (that uses 
experimental data)  with both the low-energy form dictated by chiral symmetry 
and the asymptotic behavior dictated by perturbative QCD.  
Numerical tables with our results  are available upon request.

\subsection{Determination of the $\pi \pi$ vector form factor}
The vector form factor $F_V(s)$ has been measured both directly from  
$e^+e^- \rightarrow \pi^+ \pi^-$~\cite{Amendolia:1986wj,Aloisio:2004bu,Akhmetshin:2006bx,Aubert:2009ad,Ambrosino:2010bv}
and via an isospin rotation from $\tau \rightarrow \pi^- \pi^0 \nu_\tau$~\cite{Anderson:1999ui,Fujikawa:2008ma}. 
It has also been determined by several theoretical studies \cite{Barate:1997hv,Gounaris:1968mw,Gasser:1983yg,Truong:1988zp,Kuhn:1990ad,
Gasser:1990bv,Colangelo:1996hs,Dung:1996rp,Hannah:1996tp,Hannah:1997ux,Guerrero:1997ku,Bijnens:1998fm,
Oller:2000ug,DeTroconiz:2001wt,Pich:2001pj,Ananthanarayan:2011xt,Hanhart:2012wi}.

In the spirit of Refs.~\cite{Pich:2001pj, Jamin:2006tk,Jamin:2008qg,Boito:2010me,Bernard:2011ae,Dumm:2013zh,Descotes-Genon:2013uya,Escribano:2013bca} we determine the vector form factor phenomenologically 
by fitting the invariant mass distribution of $\tau \rightarrow \pi^- \pi^0 \nu_\tau$ decays using a theoretically well-motivated parametrization.    To this end, we adapt the dispersive parametrizations introduced in Refs.~\cite{Pich:2001pj,Dumm:2013zh} mimicking what has been 
done for $K\pi$ in Refs.~\cite{Jamin:2008qg,Boito:2010me,Bernard:2011ae,BBP}. 
Note that for our purposes, the isospin-breaking  corrections can be neglected. 
A dispersion relation with three subtractions at $s=0$ is written for ln($F_V(s)$). 
This leads to the following representation for $F_V(s)$~\cite{Pich:2001pj,Dumm:2013zh} 
\be
F_V(s)=\mathrm{exp} \left[ \lambda_V'  \frac{s}{M_\pi^2} + \frac{1}{2} \left( \lambda_V''-\lambda_V'^2 \right) \left( \frac{s}{M_\pi^2}\right)^2+
\frac{s^3}{\pi} \int_{4 M_\pi^2}^\infty \frac{ds'}{s'^3}\frac{\phi_V(s')}{(s'-s-i\epsilon)} \right]~.
\label{eq:FVDR}
\ee
To fix one subtraction constant, use has been made of $F_V(s=0) \equiv 1$ required by gauge invariance. $\lambda_V' $ and $\lambda_V''$ are 
the two other subtraction constants corresponding to the slope and the curvature of the form factor.  They are determined from a fit to the data. 
$\phi_V(s)$ represents the phase of the form factor.  
In the elastic region $\left( s  \lesssim 1~\mathrm{GeV}^2\right)$, according to Watson theorem~\cite{Watson:1952ji} the phase of the form factor $\phi_V(s)$ 
is equal to the P wave $I=1$ $\pi \pi$ scattering phase shift $\delta_1^1(s)$ which is known with an excellent precision from the solutions of Roy-Steiner equations~\cite{Ananthanarayan:2000ht,CCL}. 
However for  $s  > 1~\mathrm{GeV}^2$ other channels open ($4 \pi, K\bar{K}$) and $\phi_V(s)$ is not known. 
Taking advantage of the precise measurements of the invariant mass distribution of 
$\tau \rightarrow \pi^- \pi^0 \nu_\tau$ decays~\cite{Fujikawa:2008ma}, the phase of the form factor can be modeled in terms of the three resonances found in 
this decay region and directly determined from the data. 

We write $\mathrm{tan} \phi_V(s) = \mathrm{Im} \tilde F_V(s)/ \mathrm{Re} \tilde F_V(s)$ in terms of a model for the form factor 
$\tilde F_V(s)$ that includes three resonances $\rho(770)$, $\rho'(1465)$ and $\rho''(1700)$ with two mixing parameters $\alpha'$ and $\alpha''$ 
measuring the relative weight between the resonances and $\phi'$ and $\phi''$ accounting for the corresponding interferences, see Ref.~\cite{Dumm:2013zh}:
\be
\tilde F_V(s) = \frac{\tilde M_\rho^2+\left( \alpha' e^{i\phi'}+\alpha'' e^{i\phi''}\right)s}
{\tilde M_\rho^2-s+\kappa_\rho~\mathrm{Re}\left[ A_\pi(s)+\frac{1}{2}A_K(s)\right]-i\tilde M_\rho \tilde \Gamma_\rho(s)}-
\frac{\alpha' e^{i\phi'}s}{D ( \tilde M_{\rho'}, \tilde \Gamma_{\rho'} )}-\frac{\alpha'' e^{i\phi''}s}{D ( \tilde M_{\rho''}, \tilde \Gamma_{\rho''} )}~,
\label{eq:tildeFV}
\ee
with 
\be
D(\tilde M_R, \tilde \Gamma_R)=\tilde M_R -s +\kappa_R \mathrm{Re} A_\pi (s) - i \tilde M_R \tilde \Gamma_R(s)~.
\label{eq:D}
\ee
In this equation $\tilde M_R$ and $\tilde \Gamma_{R}$ are model parameters.   $\tilde \Gamma_{R}$ and $\kappa_R$ are given by :
\be
\tilde \Gamma_{R}(s) = \tilde \Gamma_R \frac{s}{\tilde M_R^2}  \frac{\left(\sigma_\pi^3(s)+
1/2~\sigma_K^3(s)\right)}{\left(\sigma_\pi^3(\tilde M_R^2)+1/2~\sigma_K^3(\tilde M_R^2)\right)}~,
~~ \kappa_{R}(s)=\frac{\tilde \Gamma_R}{\tilde M_R}  \frac{s}{\pi \left(\sigma_\pi^3(\tilde M_R^2)+1/2~\sigma_K^3(\tilde M_R^2)\right)} ~,
\label{eq:gamkaprho}
\ee
if $R \equiv \rho$ and 
\be
\tilde \Gamma_{R}(s) = \tilde \Gamma_R \frac{s}{\tilde M_R^2}  \frac{\sigma_\pi^3(s)}{\sigma_\pi^3(\tilde M_R^2)}~,~~
 \kappa_{R}(s) = \frac{\tilde \Gamma_R}{\tilde M_R}  \frac{s}{\pi \sigma_\pi^3(\tilde M_R^2)}~,
\label{eq:gamkapR}
\ee
otherwise. 
This parametrization is guided by Resonance Chiral Theory (RChT)~\cite{Ecker:1988te,Ecker:1989yg,Cirigliano:2004ue}. 
While RChT allows one to compute the decay width $\tilde \Gamma_R$ and $\kappa_R$ for the $\rho$ resonance, Eq.~(\ref{eq:gamkaprho}), 
taking into account the $\pi \pi$ and $K \bar{K}$ intermediate states~\cite{Guerrero:1997ku}, 
this is not the case anymore for $\rho'$ and $\rho''$. Hence in Eq.~(\ref{eq:gamkapR}), generic $\tilde \Gamma_R$ and $\kappa_R$ 
as expected for a vector resonance decaying only in $\pi \pi$ has been assumed\footnote{The assumption that  $\rho'$ and $\rho''$ only  decay in $\pi \pi$ has been made. 
One could improve the model by considering other decay modes as it has been done for $K\pi$ in Ref.~\cite{BBP}.}~\cite{Jamin:2006tk}.    
In Eq.~(\ref{eq:tildeFV}), $A_\pi(s)$ and $A_K(s)$ are the $\pi \pi$ and $K \bar{K}$ loop functions in ChPT~\cite{Guerrero:1997ku,Dumm:2013zh} 
and $\sigma_\pi$ and $\sigma_K$ represents the velocity of the two particles in the centre-of-mass frame:
\begin{eqnarray}
\sigma_\pi(s) \!\! & \equiv & \!\! \sqrt{1-4 M_\pi^2/s}~\theta \left(s-4 M_\pi^2\right)~, \nonumber \\
\sigma_K(s) \!\! & \equiv & \!\! \sqrt{1-4 M_K^2/s}~\theta \left(s-4 M_K^2\right)~.
\label{eq:sigKpi}
\end{eqnarray}
Here $\theta$ denotes the Heaviside step function $\theta(x) = 1$ for $x>0$, being zero otherwise.   Note that the parameter $\kappa_R$ is defined such as $i \kappa_R~\mathrm{Im} A_\pi(s) = - i \tilde M_R \tilde \Gamma_R(s)$ with 
$\mathrm{Im} A_\pi(s) \to \mathrm{Im} [A_\pi(s)+1/2 A_\pi(s)]$ for $\rho$.  We emphasize here that $\tilde M_R$ and $\tilde \Gamma_R$ 
are model parameters and do not correspond to the physical resonance mass and width. To find them one has to find the pole of each term of 
Eq.~(\ref{eq:tildeFV}) or equivalently the zeros of their denominator Eq.~(\ref{eq:D}) on the second Riemman sheet.  

The model used to determine $\phi_V$, Eq.~(\ref{eq:tildeFV}), inspired by the Gounaris-Sakurai parametrization~\cite{Gounaris:1968mw} 
is only valid in the $\tau$ decay region and is therefore only used in Eq.~(\ref{eq:FVDR}) for $s\le s_{\mathrm{cut}} \sim m_\tau^2$. 
For the high-energy region of the dispersive integral Eq.~(\ref{eq:FVDR}) ($s> s_{\mathrm{cut}} \sim m_\tau^2$) the phase is unknown and following 
Refs.~\cite{Bernard:2011ae,Bernard:2009zm,BBP} 
we take a conservative interval between 0 and $2\pi$ centered at the asymptotic value 
of the phase of the form factor which is $\pi$. 
Indeed perturbative QCD dictates the asymptotic behavior of the form factor: it should vanish as $\mathcal{O}(1/s)$ 
up to logarithmic  corrections~\cite{Lepage:1980fj} for large values of $s$ implying that its phase should asymptotically reach $\pi$. 
The use of a three-time subtracted dispersion relation reduces the impact of our ignorance of the phase at relative high energies in Eq.~(\ref{eq:FVDR}). 
However, in order for the form factor to have the correct asymptotic behavior two sum rules have to be satisfied:
\begin{equation}
\lambda_V'^{~\mathrm{sr}} =  \frac{m_\pi^2}{\pi}  \int_{4 M_\pi^2}^{\infty} ds' \frac{\phi_V(s')}{s'^2}~,
\label{eq:sumrule1}
\end{equation}  
\begin{equation}
(\lambda_V'' - \lambda_V'^2)^{\mathrm{sr}} = \frac{2 m_\pi^4}{\pi}  \int_{4 M_\pi^2}^{\infty} ds' \frac{\phi_V(s')}{s'^3} \equiv  \alpha_{2v}^{\mathrm{sr}}~. 
\label{eq:sumrule2}
\end{equation} 
They are used to constrain the fit to the data as done for $K\pi$ in Refs.~\cite{Bernard:2011ae,BBP}. \\ \\
Twelve parameters entering $F_V(s)$, Eq.~(\ref{eq:FVDR}) are therefore determined by a fit to the data: 
\begin{itemize}
\item $\lambda_V'$ and $\lambda_V''$, the two subtraction constants   
\item $\tilde M_\rho$ and $\Gamma_\rho$, $\tilde M_{\rho'}$ and $\Gamma_{\rho'}$, $\tilde M_{\rho''}$ and $\Gamma_{\rho''}$ 
the mass and decay width of $\rho(770)$,  $\rho'(1465)$ and $\rho''(1700)$ respectively used to model the phase  
\item $\alpha'$, $\alpha''$ and their phases $\phi'$, $\phi''$ the mixing parameters between the resonances 
\end{itemize} 
The following quantity is minimized :
\be
\chi^2= \sum_{i=1}^{62}\left(\frac{\left(|F_V(s)|^2\right)_i^\mathrm{theo}-\left(|F_V(s)|^2\right)_i^\mathrm{exp}}{\sigma_{\left(|F_V(s)|^2\right)_i^\mathrm{exp}}}\right)^2 +
\left( \frac{\lambda_V'-\lambda_V'^{~\mathrm{sr}}}{\sigma_{\lambda_V'^{~\mathrm{sr}}}} \right)^2+
\left( \frac{\alpha_{2v}-\alpha_{2v}^{\mathrm{sr}}}{\sigma_{\alpha_{2v}^{\mathrm{sr}}}} \right)^2~, 
\ee
with $\left(|F_V(s)|^2\right)^\mathrm{exp}$ and its uncertainty  $\sigma_{\left(|F_V(s)|^2\right)\mathrm{exp}}$, the modulus squared of the vector 
form factor experimentally extracted from the measurement of the $\tau^- \to \pi^- \pi^0 \nu_\tau$ invariant decay 
distribution~\cite{Fujikawa:2008ma} and $F_V(s)^\mathrm{theo}$ the form factor parametrized in Eq.~(\ref{eq:FVDR}). 
In addition to the first term also minimized in previous analyses~\cite{Fujikawa:2008ma,Dumm:2013zh}, 
we impose the constraints given by the two sum rules Eqs.~(\ref{eq:sumrule1}) and 
(\ref{eq:sumrule2})\footnote{$\sigma_{\lambda_V'^{~\mathrm{sr}}}$ and $\sigma_{\alpha_{2v}^{\mathrm{sr}}}$ 
are given by the $2\pi$ band taken for the high energy phase.} to guarantee the correct asymptotic behaviour of the form factor. 
This allows us to have a description for the form factor, Eq.~(\ref{eq:FVDR}) that not only fulfills the properties of analyticity and unitarity 
but is also in agreement with perturbative QCD. This is not the 
case for the dispersive representations of Refs.~\cite{Pich:2001pj,Dumm:2013zh} and 
for the parametrization used by Belle collaboration to fit their data~\cite{Fujikawa:2008ma}. 
\begin{table}[h!]
\begin{center}
\begin{tabular}{|l||c|}
\hline
$\lambda_V' \times 10^{3} $ &  $ 36.7 \pm   0.2$  \\
$\lambda_V'' \times 10^{3} $ & $ 3.12 \pm  0.04$   \\
$\tilde M_\rho \mathrm{[MeV]} $  &   $833.9 \pm 0.6 $  \\
$ \tilde \Gamma_\rho \mathrm{[MeV]} $ & $198 \pm 1$ \\
$ \tilde M_{\rho '} \mathrm{[MeV]}$  &  $1497  \pm 7 $   \\
$\tilde \Gamma_{\rho '} \mathrm{[MeV]}$ &$785  \pm 51$  \\
$ \tilde M_{\rho ''} \mathrm{[MeV]}$  &  $1685  \pm 30 $   \\
$\tilde \Gamma_{\rho ''} \mathrm{[MeV]}$ &$800  \pm 31$  \\
$\alpha'$ & $0.173\pm 0.009$   \\
$\phi'$ & $-0.98 \pm 0.11$\\
$\alpha''$ & $0.23 \pm 0.01$\\
$\phi''$& $2.20 \pm 0.05$ \\
\hline 
\hline 
$\chi^2/d.o.f$ & $38/52$   \\
\hline 
\end{tabular}
\caption{
{\it Results for the $\pi \pi$ vector form factor parameters from a fit to $\tau \rightarrow \pi \pi \nu_\tau$ data~\cite{Fujikawa:2008ma}. 
Note that $\tilde M_R$ and $\tilde \Gamma_R$ are model parameters and do not correspond to the physical resonance mass and width.}}
\label{tab:fitFV}
\end{center}
\end{table}

\begin{figure}[ht!]
\centering
\includegraphics[width=0.7\textwidth]{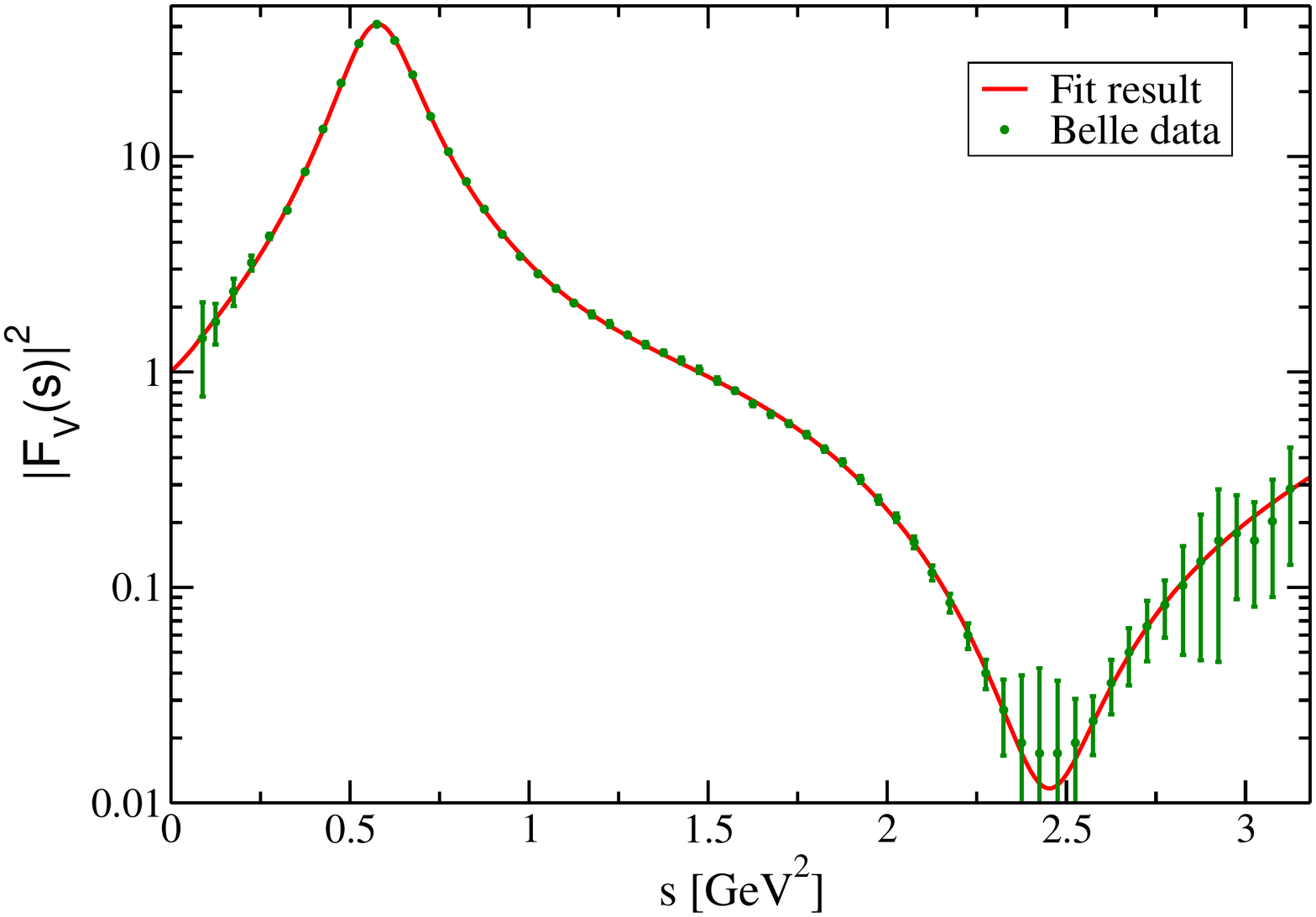}
\caption{\label{fig:Fv} \it \small  Fit result for the modulus squared of the pion vector form factor. 
The data in green are from Belle Collaboration~\cite{Fujikawa:2008ma}. 
The red line represents the result of the fit presented in Tab.~\ref{tab:fitFV}.}
\end{figure}
The result of the fit is given in Tab.~\ref{tab:fitFV}  and shown in  Fig.~\ref{fig:Fv} together with the Belle data. 
As can been seen from the figure and the $\chi^2$, the agreement with data is excellent.
Note that we have presented here a description for the form factor that represents the state-of-the-art, 
in that it relies on the fewest model assumptions and is  valid on a large energy range.
For the purposes of bounding LFV Higgs couplings,   a parametrization a la Gounaris-Sakurai
that describes well the data as the one used in Ref.~\cite{Fujikawa:2008ma} could have been sufficient.

\subsection{Determination of $\Gamma_\pi(s)$, $\Delta_\pi (s)$ and $\theta_\pi (s)$} 
The scalar form factors and $\theta_\pi (s)$ cannot be determined so directly and unambiguously from the data. 
However, they can be reconstructed from dispersive theory 
with a matching at low-energy to ChPT as pioneered in Ref.~\cite{Donoghue:1990xh}. 
As we have seen, elastic unitarity only holds at low-energy for $s \ll 1$ GeV$^2$ and in the scalar case  it is very well known that the elastic approximation 
breaks down for the $\pi \pi$ $S$-wave already at the $K\bar{K}$ threshold due to the strong inelastic coupling 
of two $S$-wave pions to $K\bar{K}$ in the region of  $f_0(980)$. In order to describe the scalar form factors 
in the kinematical region needed for $\tau \to \ell \pi \pi$,  
one has to solve a two-channel Mushkhelishvili-Omn\`es problem following 
Refs.~\cite{Donoghue:1990xh, Moussallam:1999aq} including $\pi \pi$ and $K\bar{K}$ scattering. 
As $s$ increases, a new two-body channel opens: $\eta \eta$. At some point, the $4 \pi$-channel will also become important. 
As discussed in  Refs.~\cite{Au:1986vs,Moussallam:1999aq}, below $\sqrt{s} \sim 1.4$~GeV  the inelasticity is found to 
be saturated to a good approximation by a single  channel 
$K \bar{K}$.

\subsubsection{The Mushkhelishvili-Omn\`es problem}

We  briefly recall  below the procedure presented in Ref.~\cite{Donoghue:1990xh} to solve a 
two-channel Mushkhelishvili-Omn\`es (MO) problem.   The form factors
 $F_i(s)$ ($F_1 \equiv \Gamma_\pi,  \Delta_\pi$, or $\theta_\pi$ and $\sqrt{3}/2 F_2 \equiv \Gamma_K, \Delta_K$, or $\theta_K$) 
 are analytic functions everywhere in 
 the complex plane except for a right-hand cut. 
Under the assumptions discussed above, 
the discontinuity of the form factors along the cut is determined by the two-channel 
unitarity condition: 
\be
\mathrm{Im} F_n(s) = \sum_{m=1}^2 
T^*_{nm}(s)\sigma_m(s) F_m(s)~, 
\label{eq:discFn}
\ee
where   
$T_{mn}$ represent the $T$ matrix elements which describe the scattering among  
the relevant channels  ($n =  \pi \pi, K \bar{K}$ with  $\ell=0$ and $I=0$). 
The general solution to the condition (\ref{eq:discFn})  that does not grow faster than a power of $s$ at infinity 
can be written as~\cite{Donoghue:1990xh,Muskhelishvili}: 
\be
 \begin{pmatrix}
 F_\pi (s) \\  \frac{2}{\sqrt{3}} F_K(s) 
 \end{pmatrix} 
= 
 \begin{pmatrix}
  C_1(s) & D_1(s) \\
 C_2(s) & D_2(s) 
 \end{pmatrix}
 \begin{pmatrix}
 P_F(s)  \\
Q_F(s) 
 \end{pmatrix}~,
\label{eq:solution1} 
\ee
where $P_F(s)$ and $Q_F(s)$ are polynomials and the ``canonical"  solutions $C_n(s)$, $D_n(s)$  
generalize the Omn\`es factor appearing in the solution  of the one-channel unitarity condition~\cite{Omnes:1958hv}. 

Provided that the $S$-matrix satisfies certain asymptotic conditions at large $s$ (namely that $S_{12} \to 0$ 
and   ${\rm Arg} (det (S))  \to  4 \pi$),  the solutions  $C_n(s)$ and $D_n(s)$, generically denoted by  
$X_n(s)$  behave  as $1/s$ for  $|s| \to \infty$. 
Therefore,   the $X_n(s)$   satisfy  unsubtracted dispersion relations, 
which combined with the unitarity condition (\ref{eq:discFn})  lead to   a set of coupled 
Muskhelishvili-Omn\`es singular integral equations~\cite{Muskhelishvili,Omnes:1958hv}
\be
X_n(s) =    
\sum_{m=1}^2 
\frac{1}{\pi} \int_{4 M_\pi^2}^\infty  \frac{dt}{t-s}
T^*_{nm}(t)\sigma_m(t) X_m(t)~, \qquad \qquad  X (s) =  C(s), D(s)~.  
\label{eq:Inteqs}
\ee

So in order to find a solution to the MO problem described above,  we need to specify an appropriate $T$ matrix. 
The $T$ matrix is related to the $S$ matrix by  
\be
S_{mn}=\delta_{mn} + 2i~\sqrt{\sigma_m \sigma_n}~T_{mn}~,
\label{eq:STrelation}
\ee
where the kinematical factor $\sigma_m(s)$ represents the velocity of the two particles in the centre-of-mass frame defined in Eq.~(\ref{eq:sigKpi}) with 
$\sigma_1(s) =\sigma_\pi(s)$ and $\sigma_2(s) =\sigma_K(s)$. 
In turn, the  $\ell=0$, $I=0$ projection of the  $S$ matrix is parameterized as follows
\be
S =
 \begin{pmatrix}
  \mathrm{cos} \gamma~e^{2 i \delta_\pi} & i~\mathrm{sin} \gamma~e^{i(\delta_\pi+\delta_K)}  \\
i~\mathrm{sin} \gamma~e^{i(\delta_\pi+\delta_K)} &   \mathrm{cos} \gamma~e^{2 i \delta_K}
 \end{pmatrix}~, 
\label{eq:Smatrix} 
\ee 
and therefore we need  three input functions, the inelasticity $\eta_0^0 \equiv$ cos $\gamma$,  
the $\pi \pi$ S-wave phase shift $\delta_\pi(s)$ and the $K \bar{K}$ phase shift $\delta_K(s)$. 
Up to some energy,  these inputs are determined by solving the Roy-Steiner equations for 
$\pi \pi$~\cite{Roy:1971tc, Ananthanarayan:2000ht, CCL,GarciaMartin:2011cn} and 
$K\pi$ scattering~\cite{Buettiker:2003pp}. 
Since Eq.~(\ref{eq:discFn}) is a reasonable approximation 
to the exact discontinuity only in the energy region below some cut $s_{\rm cut}   \lesssim m_\tau^2$, 
we use the following strategy: for $s < s_{\rm cut}$ we use the inputs for the two phase shifts $\delta_\pi(s)$ and $\delta_K(s)$ and the 
inelasticity $\eta_0^0(s)$ coming from a recent update of the solutions of Roy-Steiner equations~\cite{Buettiker:2003pp}
\footnote{The input values $M_\pi=139.57018$ MeV and $M_K=495.7$ MeV have been used to generate these inputs.}  provided by B.~Moussallam. 
For $s > s_{\mathrm{cut}}$,   we drive the $T$ matrix to zero consistently with unitarity, by 
forcing the three input functions to the asymptotic values   $\delta_\pi = 2 \pi$, $\delta_K = 0$, $\eta_0^0 = 1$, 
which ensure  that  the canonical solutions to the MO  problem fall off as $1/s$~\cite{Donoghue:1990xh,Moussallam:1999aq,Moussallam:2011zg}.
We have varied  $s_{\rm cut}$  in the range  $ (1.4~\mathrm{GeV})^2 - (1.8~\mathrm{GeV})^2$, 
and find that the form factors are insensitive to $s_{\rm cut}$ for $\sqrt{s} < 1.4$~GeV. 

Following Ref.~\cite{Donoghue:1990xh}, we  generate a family of solutions $\left\{ X_1 (s), X_2(s) \right\} $ of the unitary condition by iteration. 
We start with  $\left\{ X_1^{(1)}(s) = \Omega_\pi(s), X_2^{(1)} =  \lambda~\Omega_K(s) \right\} $  where $\lambda$ is a real parameter and 
$\Omega_{\pi,K}(s)$ is the Omn\`es function~\cite{Omnes:1958hv}
\be
\Omega_{\pi,K}(s) \equiv \mathrm{exp} \left[\frac{s}{\pi} \int_{4 M_\pi^2}^\infty \frac{dt}{t}\frac{\delta_{\pi,K}(t)}{(t-s)} \right]~,
\ee
solution of the one-channel unitary condition. 
We compute the iteration $(N+1)$ from iteration 
$(N)$ using Eq.~(\ref{eq:discFn})  for the imaginary part and Eq.~(\ref{eq:Inteqs}) for the real part.  
The problem admits two independent solutions~\cite{Muskhelishvili} that are 
linear combinations of the family of solutions labelled by the parameter $\lambda$ we have found. They are chosen such that~\cite{Donoghue:1990xh}
\be
C_n(s)|_{s=0}=\delta_{n1},~~D_n(s)|_{s=0}=\delta_{n2}~.
\label{eq:CnDn}
\ee

\subsubsection{Fixing the subtraction constants with chiral symmetry}

The form factors  $F_{\pi,K}(s)$  (with $F \in \{  \Gamma, \Delta, \theta \} $) 
are obtained from (\ref{eq:solution1})  once 
the polynomials $P_F(s)$ and $Q_F(s)$ are given. 
The polynomials can be determined by matching the form factors to their 
ChPT expressions at low energy~\cite{Donoghue:1990xh}, as summarized below.

For $\Gamma_{\pi,K} (s)$ and  $\Delta_{\pi,K} (s)$, 
the requirement that the form factors 
behave as $\mathcal{O} (1/s)$ for large values of $s$ fixes the polynomials to be constants. 
The polynomials are then determined by the values of the form factors at $s=0$, 
which are related to the response of the pseudoscalar masses to changes in the quark masses (Feynman-Hellmann theorem): 
\be
\Gamma_{P}(0) = \left( m_u \frac{\partial}{\partial m_u} + m_d \frac{\partial}{\partial m_d}   \right) M_P^2~,
~~\Delta_{P} (0)= \left(  m_s \frac{\partial}{\partial m_s}  \right) M_P^2~.
\label{eq:FeyHell}
\ee
The above relations imply~\cite{Donoghue:1990xh}:
\begin{eqnarray}
P_\Gamma (s)  & = &  \Gamma_\pi (0) = M_\pi^2  + \cdots  \\
Q_\Gamma (s)  & = &  \frac{2}{ \sqrt{3}}  \Gamma_K (0) = \frac{1}{\sqrt{3}}   M_\pi^2  + \cdots  \\
P_\Delta  (s)  & = &  \Delta_\pi (0) = 0  + \cdots  \\
Q_\Delta (s)  & = &  \frac{2}{ \sqrt{3}} \Delta_K (0) = \frac{2}{\sqrt{3}}  \left( M_K^2 -   \frac{1}{2} M_\pi^2 \right)  + \cdots  ~, 
\end{eqnarray}
where in the second equality above we have given the leading chiral order result and the 
dots  represent higher order corrections. 
 For the pion form factors, we  neglect the higher order chiral corrections expected to be of order  $M_\pi^2/(4 \pi F_\pi)^2$. 
However, for the kaon form factors the chiral corrections are not a priori negligible. 
They can be calculated within $SU(3)$ ChPT in terms of  low-energy constants  estimated from lattice  QCD~\cite{Colangelo:2010et}. 
These corrections have also been recently evaluated from lattice data in the framework of Resumed ChPT~\cite{Bernard:2012fw}. 
We take the ranges  $\Gamma_K(0) = (0.5 \pm 0.1)~M_\pi^2$, $\Delta_K(0) = 1^{+0.15}_{-0.05}\left( M_K^2 - 1/2 M_\pi^2\right)$~\cite{Daub:2012mu} 
that encompass the  recent estimates.

For $\theta_{\pi,K} (s)$   requiring that $P_\theta (s)$ and $Q_\theta (s)$ be constant 
(to enforce $\theta_{\pi, K} \sim {\cal O}  (1/s)$  asymptotically) 
is not consistent with the behavior in the chiral regime~\cite{Donoghue:1990xh}. 
This is a signal that the unsubtracted dispersion relation for these form factors is 
not saturated by the two states considered in the analysis. 
Relaxing the requirement on the asymptotic behavior and matching to  ChPT 
expressions implies 
\begin{eqnarray}
P_\theta (s)  & = &   2  M_\pi^2  +     \left(  
\dot \theta_\pi  - 2 M_\pi^2 \dot C_1 - \frac{4 M_K^2}{\sqrt{3}} \dot D_1
\right) \,  s   
\\
Q_\theta (s)  & = &  \frac{4}{ \sqrt{3}}  M_K^2    + 
\frac{2}{\sqrt{3}} 
\left(\dot \theta_K  - \sqrt{3} M_\pi^2 \dot C_2  - 2M_K^2 \dot D_2
\right) \, s  ~ , 
\end{eqnarray} 
where  $\dot{f} \equiv (d f /ds) (s=0)$. 
$\dot{\theta}_\pi = 1$ up to small chiral $SU(2)$ corrections. 
At leading chiral  order  $\dot \theta_K = 1$. 
An alternative  procedure to estimate $\dot \theta_K$, 
taking into account chiral $SU(3)$ corrections, 
has been given in Ref.~\cite{Donoghue:1990xh}. 
The approach is 
based on writing an unsubtracted dispersion relation for 
$\theta_K(s) -\theta_\pi(s)$: this leads to  $\dot \theta_K = 1.15 - 1.18$, 
depending on the value of  $s_{\rm cut}$ adopted. 
Based on this,  in what follows we  use  the range $\dot \theta_K = 1.15 \pm 0.1$.

\subsubsection{Results} 
\begin{figure}[ht!]
\centering
\includegraphics[width=0.49\textwidth]{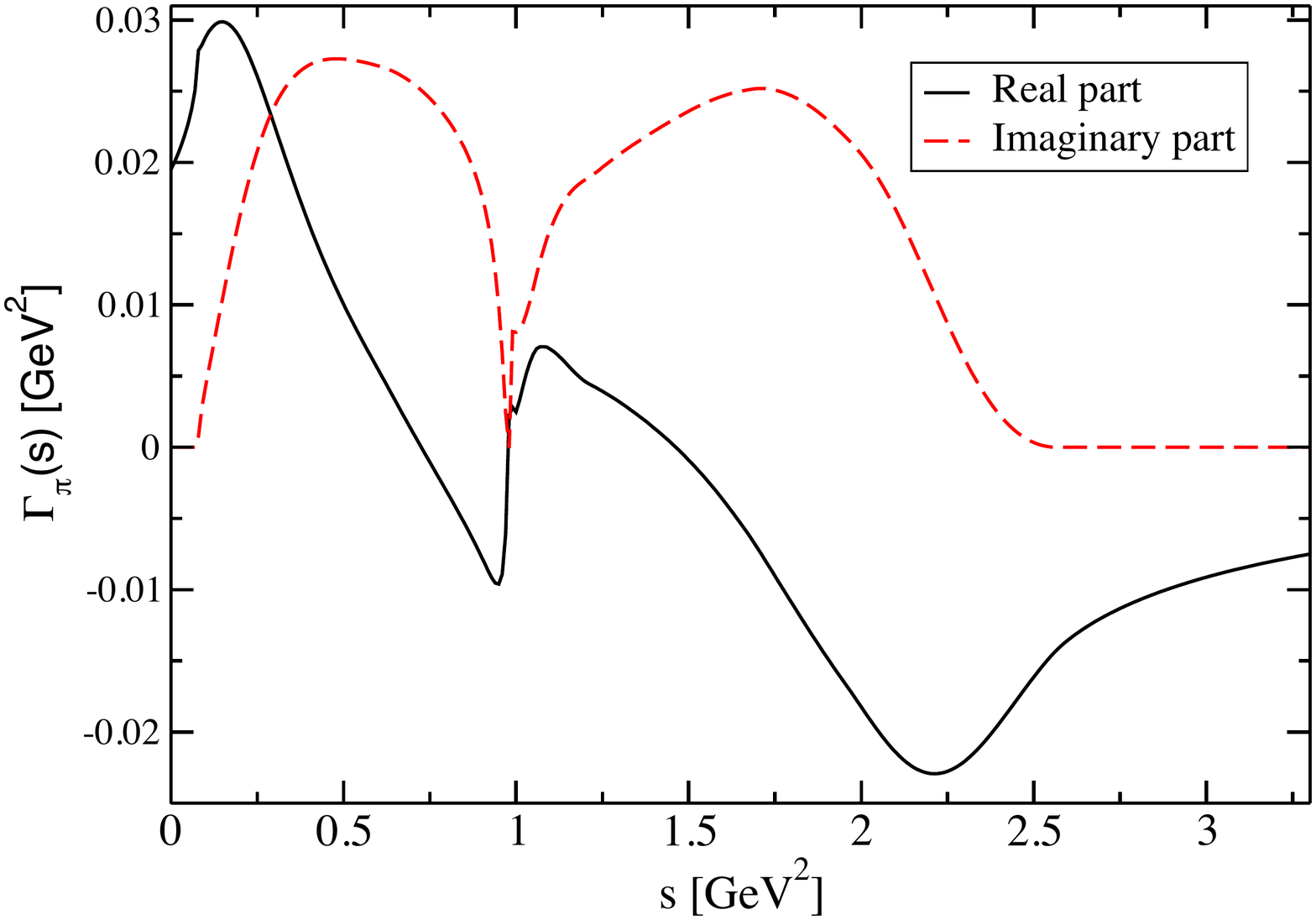}
\includegraphics[width=0.49\textwidth]{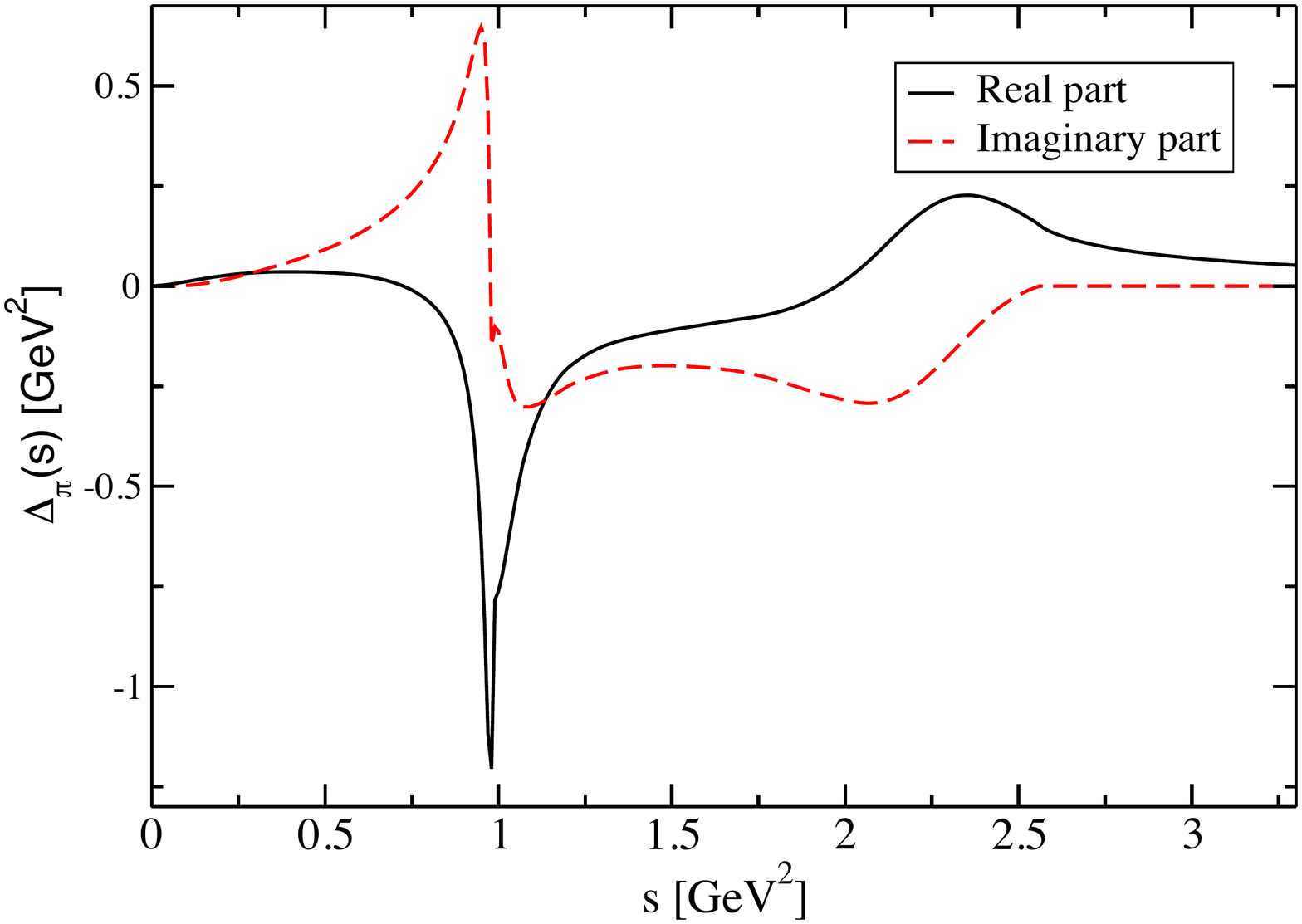}
\includegraphics[width=0.49\textwidth]{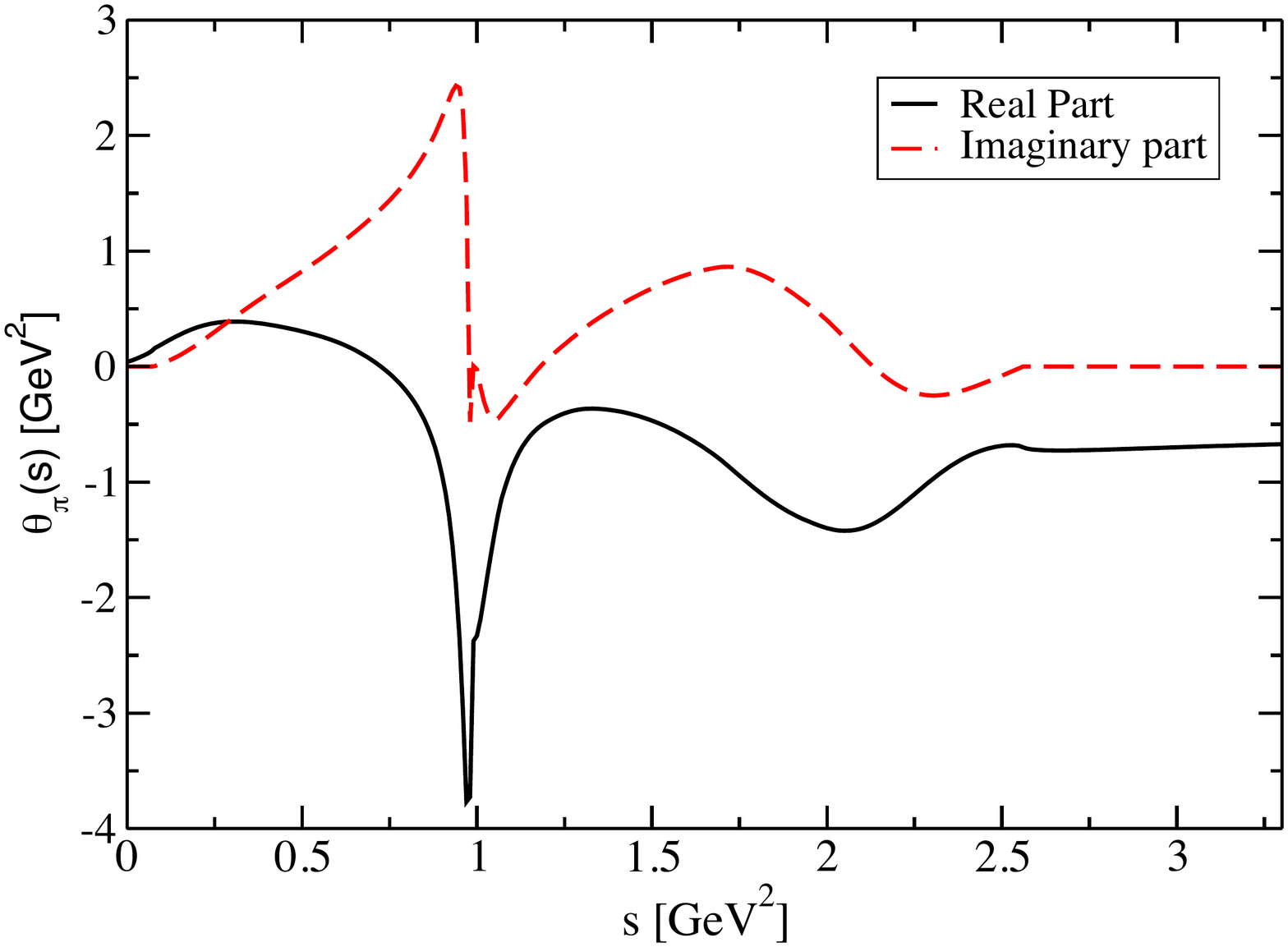}
\caption{\label{fig:ScalarsFF} \it \small  The  form factors $\Gamma_\pi(s)$, $\Delta_\pi(s)$ and $\theta_\pi(s)$ defined in Eq.~(\ref{eq:scalFFdef}) 
as determined by solving the two-channel unitarity condition  and then by matching to $ChPT$, see text for details. 
The black solid line represents their real part and the red dashed-dotted red line stands for their imaginary part.
This plot is generated using  $s_{\rm cut} =  (1.4 \,  \rm{GeV})^2$  and central values for the matching coefficients. }
\end{figure}
Using the two sets of solutions $\{C_1(s), C_2(s)\}$ and $\{D_1(s), D_2(s)\}$  and the polynomials determined in the last subsection 
we can construct the three form factors $\Gamma_\pi(s)$, $\Delta_\pi(s)$ and $\theta_\pi(s)$ from Eq.~(\ref{eq:solution1}). 
They are shown in Fig.~\ref{fig:ScalarsFF}  using  $s_{\rm cut} =  (1.4 \,  \rm{GeV})^2$  and 
central values for the matching coefficients. 
For  $\sqrt{s} < 1.4$~GeV, the form factor are relatively insensitive to the choice of  $s_{\rm cut}$: 
the dependence on  $s_{\rm cut}$ induces variations of the  $\tau \to \ell \pi \pi$  phase space integrals at the $< 15\%$ level. 
Likewise, varying the matching  polynomials  in the ranges specified in the previous subsection 
leads to changes in the phase space integrals at the level of $10\%$.  

Note that a similar approach to describe  $\Gamma_\pi(s)$, $\Delta_\pi(s)$
to study lepton flavour violating effects within  R-parity violating supersymmetry  
has been implemented in Ref.~\cite{Daub:2012mu} improving the hadronic treatment used in Ref.~\cite{Arganda:2008jj}. 
Compared to previous work, we include the effective Higgs-gluon interaction induced by the Higgs coupling to heavy quarks.   The influence of heavy quarks  is not small and provides in general the dominant contribution to low-energy hadronic transitions mediated by scalar bosons associated to the mechanism of EWSB, the Higgs coupling to light quarks being mass suppressed.   This well known  fact  has been discussed for example within the context of Higgs-nucleons interactions~\cite{Shifman:1978zn} and for the decay of a very light Higgs into two pions, $H \rightarrow \pi \pi$~\cite{Donoghue:1990xh}.  Here we provide for the first time an adequate description of this effect for Higgs mediated semileptonic $\tau \rightarrow \ell \pi \pi$ decays.

\section{Phenomenology}
\label{sec:Pheno}
Having developed the necessary form factors to describe Higgs mediated LFV $\tau \rightarrow \ell \pi \pi$ decays in the previous section,  we proceed to analyze the role of semileptonic $\tau$ decays to probe for LFV effects in the scalar sector.    We discuss the robustness of the bounds obtained compared with previous treatments in the literature that rely on LO-ChPT predictions.  We also analyze the connection between semileptonic $\tau$ decays and other LFV $\tau$ decays as well as with the search for LFV Higgs decays at the LHC.  The phenomenology of a CP-odd Higgs boson with LFV couplings is discussed with a similar spirit.  A general two-Higgs-doublet model is introduced to motivate the discussion of LFV effects in the scalar sector, however all the results in this section are expressed using the Lagrangian in Eq.~(\ref{laggen}) 
and can therefore be interpreted within other new physics scenarios.

\subsection{
2HDM and beyond
}
\label{sec:Framework}

Two-Higgs-doublet models (2HDM) provide a specific gauge-invariant framework where lepton flavor violating effects 
encoded in Eq.~(\ref{laggen})  can occur, due to both CP-even and CP-odd Higgs bosons at tree-level. 
In the Higgs basis, where only one scalar doublet acquires a vev, one can write~\cite{Gunion:1989we}
\begin{align}
\Phi_1=  \begin{pmatrix}
     G^+ \\
     \frac{1}{\sqrt{2}}(v+ S_1 + i G^0)
 \end{pmatrix} \, ,\qquad
\Phi_2= \begin{pmatrix}
      H^+ \\
     \frac{1}{\sqrt{2}}(S_2 + i S_3)
 \end{pmatrix} \,.
\end{align}
The fields $ S_{1,2}$ are CP-even while $S_3$ is a CP-odd field.   The most general Yukawa Lagrangian of the 2HDM is given by
\begin{align}
\mathcal{L}_{Y} \;= &\;   - \frac{\sqrt{2}}{v}  \Bigl\{   \bar L_L^{\prime} \, \left(   M_{\ell}^{\prime} \, \Phi_1  + \Pi_{\ell}^{\prime} \, \Phi_2  \right) \, \ell_R^{\prime}    \nonumber \\
&+   \bar Q_L^{\prime} \, \left(   M_{d}^{\prime} \, \Phi_1  + \Pi_{d}^{\prime} \, \Phi_2  \right) \, d_R^{\prime} +   \bar Q_L^{\prime} \, \left(   M_{u}^{\prime} \, \tilde{\Phi}_1  + \Pi_{u}^{\prime} \, \tilde{\Phi}_2  \right) \, u_R^{\prime}    \Bigr\} +     \mathrm{h.c.}   \,, 
\end{align}
where $\bar L_L^{\prime}  = \left( \bar \nu_L^{\prime} , \bar \ell_L^{\prime}  \right)$ is the left-handed lepton doublet, $Q_L^{\prime}  = (u_L^{\prime} , d_L^{\prime} )$ is the left-handed quark doublet and $\ell_R^{\prime}, u_R^{\prime} , d_R^{\prime} $ are right-handed SU(2) singlets in flavor space. The primes over the fields and couplings denote that we are still in an arbitrary weak basis, the usual notation $\tilde \Phi = i \tau_2 \Phi^*$ is used.    The matrices $M_{f=u,d,\ell}^{\prime}$ represent the non-diagonal fermion mass matrices while $\Pi_{f=u,d,\ell}^{\prime}$ are arbitrary complex matrices in flavor space parametrizing the Yukawa couplings of the second Higgs doublet $\Phi_2$.   

After EWSB the neutral mass eigenstates $\varphi_k = \{h, H, A\}$ are obtained via an orthogonal transformation, $\varphi_k = \mathcal{R}_{km} \, S_m$, that diagonalizes the mass matrix of the scalar fields ($M_h \leq M_H$ in our conventions).    The Higgs couplings to vector bosons $(VV= W^+W^-, ZZ)$ are given by $g_{\varphi_k VV} = \mathcal{R}_{k1} \,  g^{\mathrm{SM}} _{hVV} $, where $g^{\mathrm{SM}} _{hVV}$ represents the SM gauge coupling, while  the interactions of the physical scalars $\{h,H,A, H^{\pm}\}$ with fermions are described by
\beqn\label{lagrangianl}
 \mathcal L_{Y} &= &  - \, \sum_{\varphi_k, f= u, d, \ell} \,   \varphi_k \, \bar{f}\,  Y_{f}^{\varphi_k} \,  P_R \, f  \nonumber \\
\; && - \frac{\sqrt{2}}{v}    \, H^{+}   \, \left\{    \bar u \left[   V \, \Pi_{d} \,P_R - \Pi_{u}^{\dag}\,  V \, P_L     \right] d \,   + \bar \nu \, \Pi_{\ell} \,P_R \, \ell \,    \right\}  +    \;\mathrm{h.c.}   \,.
\eeqn
Here $V$ represents the CKM matrix, $P_{L,R} = (1  \mp \gamma_5)/2$ are the usual chirality projectors and
\begin{align} 
v \,Y_{d, \ell}^{\varphi_k} &=  M_{d, \ell} \, \mathcal{R}_{k1} + \Pi_{d, \ell} \, \left( \mathcal{R}_{k2} + i \,\mathcal{R}_{k3}  \right)   \,, \nonumber \\
v \,Y_{u}^{\varphi_k} &=    M_{u} \, \mathcal{R}_{k1} + \Pi_{u} \, \left( \mathcal{R}_{k2} -  i \,\mathcal{R}_{k3}  \right) \,,
\end{align}
 where $M_{f=u,d,l}$ are the diagonal fermion mass matrices and $\Pi_{f=u,d,l}$ remain arbitrary complex matrices in the most general case, giving rise to tree-level FCNCs.   In the Type III 2HDM~\cite{Cheng:1987rs} for example, one assumes a Yukawa structure of the form
$(\Pi_{f})_{ij}\; = \;  \lambda_{ij}  \,\sqrt{m_i m_j} $, where the dimensionless parameters $\lambda_{ij}$ are in principle of $\mathcal{O}(1)$.  Note that due to the orthogonality of the mixing matrix $\mathcal{R}$,  the scalar couplings satisfy the following sum rule
\be
\sum_k  (Y_{f}^{\varphi_k})_{ij} \, \mathcal{R}_{k1} \; = \; 0  \,  \qquad \text{for}~i \neq j  \,.
\ee
In the CP-conserving limit of the 2HDM, the matrices $\Pi_{f}$ are real and the physical fields $\varphi_k$ have definite CP-quantum numbers.  The field $A = S_3$ is CP-odd while the CP-even fields $h, H$ are a mixture of $S_{1,2}$, 
\bel{eq:CPC_mixing}
\left(\ba h\\ H  \\ A \ea\right)\; = \;
\left[\batt \cos{\tilde\alpha} & \sin{\tilde\alpha} &0   \\ -\sin{\tilde\alpha} & \cos{\tilde\alpha} & 0  \\  0 & 0 & 1   \\    \ea\right]\;
\left(\ba S_1\\ S_2  \\ S_3\ea\right) \,,
\ee
where $\tilde \alpha$ is a real rotation angle which can be expressed in terms of the Higgs potential parameters\footnote{See Ref.~\cite{Celis:2013rcs} for a discussion of the 2HDM scalar potential in the CP-conserving limit as well as in the CP-violating case.}.   Comparing with the generic Lagrangian presented in Eq.~\eqref{laggen} we derive the following matching for the CP-conserving limit,
\begin{align} \label{p1}
Y_{ij}^{h} \;&=\;  \frac{(M_{f})_{ij}}{v} \cos \tilde \alpha + \frac{ (\Pi_{f})_{ij} }{v}  \, \sin \tilde \alpha   \,, \nonumber \\
Y_{ij}^{H} \;&=\; -  \frac{(M_{f})_{ij}}{v} \sin \tilde \alpha + \frac{(\Pi_{f})_{ij}}{v}  \,\, \cos \tilde \alpha   \,, \nonumber \\
Y_{ij}^{A} \; &= \;  \pm i \, \frac{(\Pi_{f})_{ij}}{v}   \,  \,.
\end{align}
 The plus sign in the expression for $Y_{ij}^{A}$ is for $f= d, \ell$ while the minus sign is for $f =u$.    In this limit $g_{hVV} = \cos \tilde \alpha\, g^{\mathrm{SM}} _{hVV}$, $g_{HVV} = - \sin \tilde \alpha\, g^{\mathrm{SM}} _{hVV}$ and $g_{AVV} = 0$. We can see that certain relations between the LFV scalar couplings arise in this case.   The fermionic couplings of the lightest CP-even Higgs are flavor conserving in the limit $g_{hVV} = g^{\mathrm{SM}} _{hVV} $, and, in general these are suppressed by the factor $\sin \tilde \alpha$. Flavor-changing couplings of the CP-odd Higgs on the other hand do not receive such suppression.    Considering the 2HDM to be a low-energy effective theory, the effect of heavy degrees of freedom contained in a UV completion will in general introduce corrections to Eq.~\eqref{p1}, spoiling these specific relations.  The effective Lagrangian of dimension-six for example contains the following terms that modify the Yukawa structure of the 2HDM,
 \be \label{mlfveq}
 \mathcal{L}_{d=6} \supset  \sum_{p,r,s} \frac{1}{\Lambda^2}    \left( C^{prs}_{\ell} \, \Phi_p^{\dag} \Phi_r    \, ( \bar L_L  \, \ell_R \,  \Phi_s ) \, + \mathrm{h.c.}  \right)    + \cdots \,,
 \ee
 where $\Lambda$ represents the scale of new physics (beyond the 2HDM) and $C^{prs}_{\ell} $ are arbitrary coefficients with the role of dimensionless low-energy constants that encode the high-energy dynamics.   Corrections of the type~\eqref{mlfveq} are in general very small since these are suppressed by inverse powers of the high energy scale $\Lambda$.  For the LFV couplings however these corrections can become relevant since it is possible that these couplings vanish or are very small in the low-energy theory. 

\subsection{A CP-even Higgs with LFV couplings}
\label{sec:CPeven}

The phenomenology of a CP-even Higgs at $125$ GeV with LFV couplings has been analyzed recently in Refs.~\cite{Blankenburg:2012ex,Harnik:2012pb,Davidson:2012ds}.  It has been noticed that large branching fractions for the decay $h \rightarrow \tau \mu$ are possible ($\mathrm{BR}(h \rightarrow \tau \mu) \lesssim 0.1$) while being compatible with present low-energy constraints from $\tau \rightarrow \mu \gamma$ and $\tau \rightarrow \mu \mu \mu$.   Higgs decays into a $\tau- e$ pair can also have large branching fractions consistent with low-energy flavor constraints while $h \rightarrow e \mu $ is already strongly bounded by $\mu \rightarrow e \gamma$~\cite{Blankenburg:2012ex}. 
In Ref.~\cite{Davidson:2012ds} it has been estimated that the LHC can set an upper bound $\mathrm{BR}(h \rightarrow \tau \mu) \lesssim  4.5 \times 10^{-3}$ with $20$~fb$^{-1}$ of data with Higgs production occurring mainly through the dominant gluon fusion mode.  

 The strongest low-energy constraint on possible $\tau$-$\ell$ LFV couplings of the $125$~GeV Higgs is obtained from the process $\tau \rightarrow \ell \gamma$.  This decay occurs at the loop-level and receives dominant contributions from two-loop diagrams of the Barr-Zee type since the one-loop diagrams are chirality suppressed~\cite{Chang:1993kw}.   The LFV radiative $\tau$ decay however is not directly related to the process $pp (gg) \rightarrow h \rightarrow \tau \ell$ observable at the LHC.  Indeed, heavy degrees of freedom belonging to the UV completion of the theory or additional scalars from an extended Higgs sector could contribute also to the effective dipole operator~$(\bar \ell  \sigma^{\mu \nu} P_{L,R} \tau ) F_{\mu \nu}$, making the bound extracted very model dependent. For example, 
in the simple scenario of a 2HDM, the additional neutral Higgs bosons $A$ and $H$ 
generate interfering contributions through diagrams similar to the ones involving $h$.
These effects  cannot be neglected in general~\cite{Chang:1993kw,Davidson:2010xv}.

It is therefore important to consider processes that can give a more reliable bound on the LFV couplings of the $125$~GeV Higgs and, also, which are more directly connected with the observables measured at the LHC.  Besides light quark exchange, the same effective vertex of the Higgs to gluons responsible for the production of the Higgs via gluon fusion at the LHC, would also contribute to the semileptonic decay $\tau \rightarrow \ell \pi \pi$  ($\pi \pi = \pi^+ \pi^-, \pi^0 \pi^0$), see Fig.~\ref{taumupipivsLHC}.  The energy scale of these  processes however are completely different and in opposite domains of QCD: the LHC process $gg \rightarrow h$ occurs in the perturbative domain of QCD, while the $\tau$ decays takes place at an intermediate scale where non-perturbative QCD effects play a crucial role (one has to consider the matrix element $\langle \pi \pi | G_{\mu \nu}^{a}  G^{\mu \nu}_{a} |Ê0 \rangle$).     At the energy scale relevant for $\tau$ decays, the effective Lagrangian describing the interactions of the Higgs with light-quarks and gluons is given by~\cite{Shifman:1978zn}
\begin{equation}  \label{effLagg}
 \mathcal L_{eff}^{h}  \simeq - \frac{h}{v}  \,    \left(   \, \sum_{q=u,d,s}  \,  y_{q}^{h} \, m_{q} \, \bar q    \,q   -   \sum_{q=c,b,t}  \,   \dfrac{ \alpha_s \,}{12 \pi} \,  y_q^{h}\,   G_{\mu \nu}^{a}  G^{\mu \nu}_{a} \right)     \,.
\end{equation}
Neglecting $m_\ell$, the differential decay width for the decay $\tau \rightarrow \ell \pi \pi$ mediated by the CP-even Higgs $h$ can be written 
in terms of the two-pion invariant mass $\sqrt{s}$ ( $s = (p_{\pi^+} + p_{\pi^-})^2 = (p_{\tau} - p_{\ell})^2$)  as 
\begin{align}
\label{diffHiggs}
\frac{d \Gamma(\tau \rightarrow \ell \pi^+ \pi^-)_{ \text{Higgs}} }{d \sqrt{s}} &\;=\;   
\frac{ (m_{\tau}^2 -s)^2    \left( s - 4 m_\pi^2\right)^{1/2}  }{256\,\pi^3\,  m_{\tau}^3}  \cdot 
\frac{  |Y_{\tau \ell}^{h}|^2 + |Y_{\ell \tau}^{h}|^2 }{ M_h^4 \,v^2} 
\nonumber \\ 
&  \ \times   \Big|   \mathcal{K}_{\Delta}  \Delta_{\pi}(s) +    \mathcal{K}_{\Gamma}  \Gamma_{\pi}(s) +    \mathcal{K}_{\theta}  \theta_{\pi}(s)   \Big|^2         \,, \nonumber \\
 \frac{d \Gamma(\tau \rightarrow \ell \pi^0 \pi^0)_{\text{Higgs}} }{d\sqrt{s}}  &\;=\;   \frac{1}{2}  \frac{d \Gamma(\tau \rightarrow \mu \pi^+ \pi^-)_{\text{Higgs}} }{d\sqrt{s}} \ , 
\end{align}
where 
\begin{align}
\label{eq:sc}
 \mathcal{K}_{\theta}  =  \frac{2}{27} \sum_{q=c,b,t} y_q^h    \,,   \qquad
 \mathcal{K}_{\Delta} =   y_s^h -   \mathcal{K}_{\theta}       \,, \qquad 
 \mathcal{K}_{\Gamma} =  \frac{  m_u y_u^h + m_d y_d^h  }{m_u+m_d}    - \mathcal{K}_{\theta}     \,.
\end{align} 
The form factors $\Gamma_{\pi}(s)$, $\Delta_{\pi}(s)$, and $\theta_\pi (s)$  parametrize the hadronic matrix elements of the scalar-currents  
and the gluonic operators (see  Eqs.~(\ref{eq:trace}) and (\ref{eq:scalFFdef})). 

At the loop-level, a LFV Higgs also generates an effective dipole operator 
\be \label{dipole}
\mathcal{L}_{\mathrm{eff}} \; = \; c_L \, Q_{L\gamma} + c_R \, Q_{R \gamma} + \mathrm{h.c.} \,,
\ee
with
\be
Q_{L\gamma,R\gamma} \; = \; \frac{e}{8 \pi^2} m_{\tau} \, \left(  \bar \ell \, \sigma^{\alpha \beta} \, P_{L,R} \, \tau  \right) F_{\alpha \beta}\,.
\ee
For the evaluation of the Wilson coefficients $c_{L,R}$ we consider one-and two-loop contributions calculated in Ref.~\cite{Chang:1993kw} and recently discussed in Ref.~\cite{Harnik:2012pb}.    
The effective dipole operator gives rise to $\tau \rightarrow \ell \pi^+ \pi^-$ via photon exchange, the associated differential decay width is given by 
(neglecting small lepton mass effects)
\begin{align} \label{diffDipole}
\frac{d \Gamma(\tau \rightarrow \ell \pi^+ \pi^-)_{\text{photon}} }{ d\sqrt{s} } \; &= \;    \frac{ \alpha^2 (|c_L|^2 + |c_R|^2)    }{  768  \pi^5  \, m_{\tau} } 
\cdot  \frac{(s - 4 m_\pi^2)^{3/2}  \,  (m_\tau^2 - s)^2  \,  (s + 2 m_\tau^2)   \  |F_{V}(s)|^2 }{s^2} \,,
 \nonumber \\
\frac{d \Gamma(\tau \rightarrow \ell \pi^0 \pi^0)_{\text{photon}} }{ d\sqrt{s}  } \; &= \; 0   \,,
\end{align}
where $F_{V}(s)$ is the pion vector form factor defined in Eq.~(\ref{FVdef}).

\begin{figure}[t!]
\centering
\includegraphics[width=0.6\textwidth]{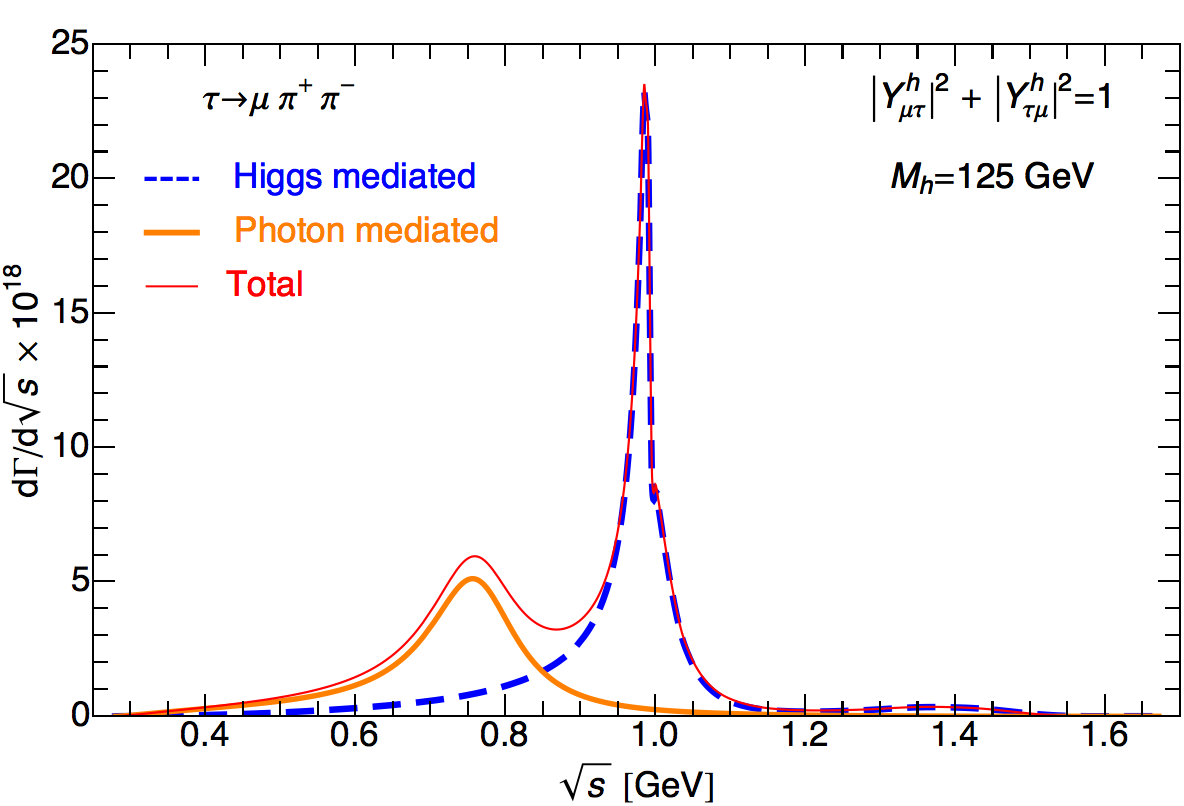}
\caption{\label{fig:spectra} \it \small  
$\tau \to \mu \pi^+ \pi^-$ differential decay rate vs the di-pion invariant mass $\sqrt{s}$: 
dipole contribution (thick solid orange line),  Higgs-mediated contribution (dashed blue line), 
and total rate (thin solid red line). 
The diagonal couplings of the Higgs are fixed to their SM values. 
} 
\end{figure}

The Higgs (Eq.~\eqref{diffHiggs}) and photon exchange (Eq.~\eqref{diffDipole}) contributions do not interfere so that $\Gamma_{\text{total}} = \Gamma_{\text{Higgs}} + \Gamma_{\text{photon}} $. While the $\tau \rightarrow \ell \pi^+ \pi^-$ channel can be mediated by photon exchange, the $\tau \rightarrow \ell \pi^0 \pi^0$ mode does not receive any contributions from intermediate photons due to the Bose statistics of the hadronic final state.  The $\tau \rightarrow \mu \pi^0 \pi^0$ decays therefore do not receive any contributions from the effective dipole operator $(\bar \ell \sigma^{\mu \nu} P_{L,R} \tau) F_{\mu \nu}$ and isolates the CP-even Higgs exchange contribution (a CP-odd Higgs cannot mediate $\tau \rightarrow \mu \pi \pi$ decays due to spin-parity conservation).   

In Figure.~\ref{fig:spectra} we plot the $\tau \to \mu \pi^+ \pi^-$
differential decay rate in the di-pion invariant mass $\sqrt{s}$,  using  the benchmark input values 
$M_h= 125$~GeV, $y_{f}^{h} =1$, and $  |Y_{\tau \ell}^{h}|^2 + |Y_{\ell \tau}^{h}|^2 = 1$. 
The dipole contribution is characterized by the  $\rho$ resonance peak,   while the Higgs-mediated contribution 
(dominated by $\Delta_\pi (s)$ and $\theta_\pi (s)$) 
is characterized by  the sharp $f_0(980)$ peak.  
Clearly,  a measurement of the spectrum  would greatly help disentangling the underlying  LFV mechanism. 

\begin{figure}[t!]
\centering
\includegraphics[width=7.5cm,height=6.cm]{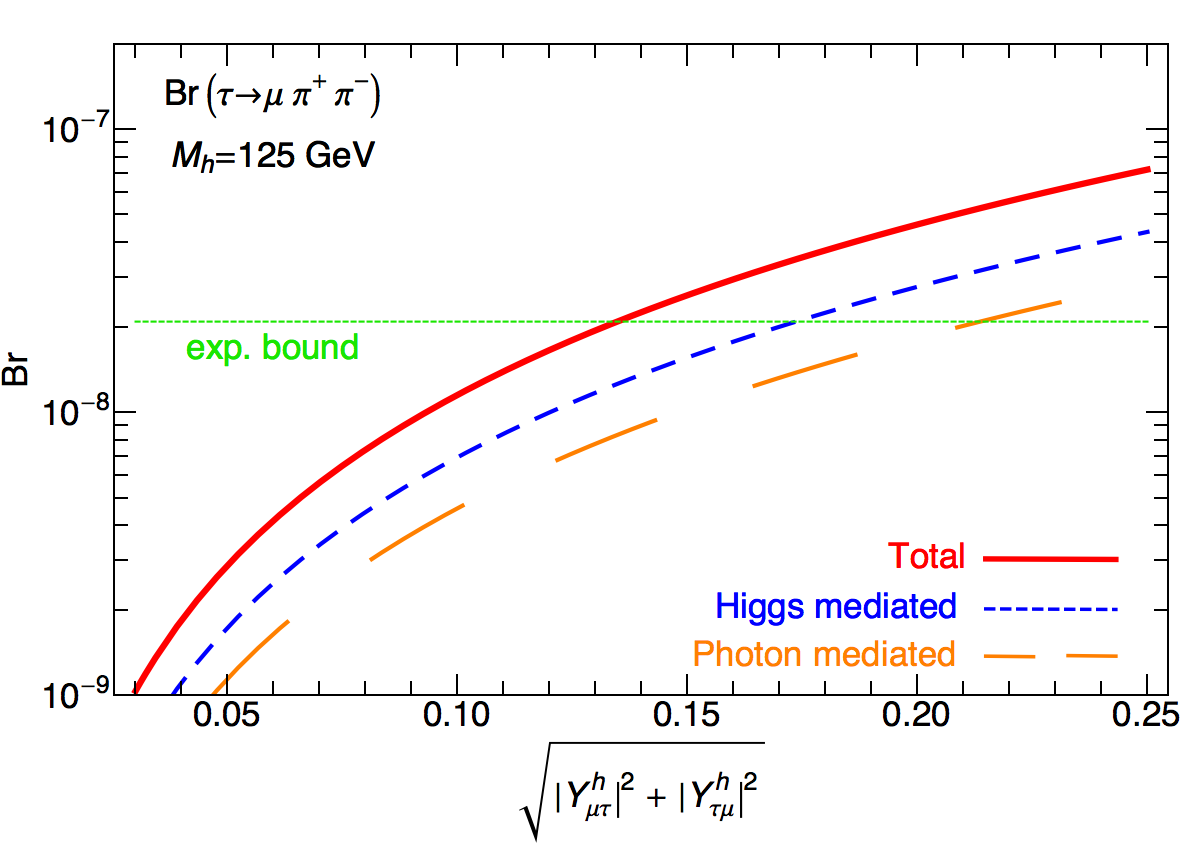}
~
\includegraphics[width=7.5cm,height=6cm]{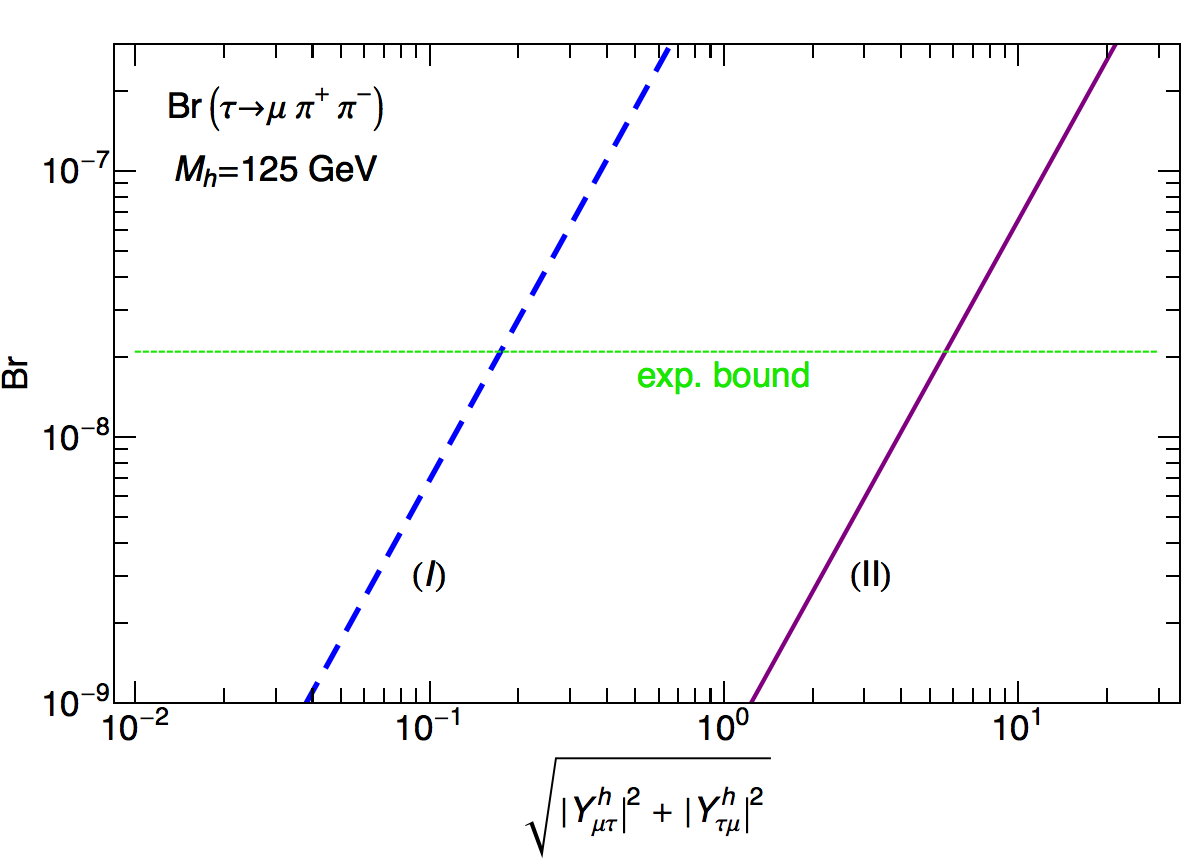}
\caption{\label{taumupipitotal} \it \small  Left-plot: Branching fraction for $\tau \rightarrow  \mu \pi^+ \pi^-$ as a function of  $\sqrt{ |Y^{h}_{\mu \tau}|^2 + |Y^{h}_{\tau \mu}|^2  }$ for $M_{h} = 125$~GeV and SM-like diagonal couplings (continuous red line).  The effective dipole contribution is shown in orange (long-dashed) while the Higgs exchange contribution is shown in blue (short-dashed). The present experimental upper bound is shown in green (horizontal dashed line).  Right-plot:  Higgs mediated contribution to the branching ratio considering: (I) our prediction for the form factors $\{\Gamma_{\pi}(s), \Delta_{\pi}(s), \theta_{\pi}(s) \}$ using ChPT and dispersion relations,  (II) estimate usually found in the literature considering only the scalar form factor ${\Gamma_{\pi} (s)}$ in LO-ChPT. } 
\end{figure}

The branching fraction for $\tau \rightarrow \mu \pi^+ \pi^-$ is shown in Fig. \ref{taumupipitotal} (left-panel) as a function of the combination of LFV couplings $\sqrt{|Y_{\tau \mu}^{h}|^2 + |Y_{\mu \tau}^{h}|^2}$, the mass of the Higgs is fixed at $M_h= 125$~GeV and the diagonal fermionic Higgs couplings are taken to be SM-like ($y_{f}^{h} =1$).  
We use here
the form factors  determined in Sect.~\ref{sec::ffs}. 
The short-dashed (blue) curve shows the Higgs mediated contribution Eq.~\eqref{diffHiggs} while the long-dashed (orange) curve shows the photon mediated one Eq.~\eqref{diffDipole}.  The total branching fraction is shown as a continuous (red) line.  In Fig.~\ref{taumupipitotal} (right-panel) we compare our prediction for the Higgs mediated contribution to the one usually  considered in the literature~\cite{Black:2002wh,Kanemura:2005hr,Arganda:2008jj}, which is based on leading order ChPT predictions for $\Gamma_{\pi}(s)$, see Sect.~\ref{sec:discu} for a detailed discussion.

In Tab.~\ref{tab:tableCPeven} we show the bounds that different LFV $\tau$ decays put on the combination of LFV couplings $\sqrt{ |Y^{h}_{\mu \tau}|^2 + |Y^{h}_{\tau \mu}|^2}$, assuming SM-like diagonal Yukawa couplings ($y_{f}^{h}=1$).  Similar bounds for $\tau - e$ transitions are shown in Tab.~\ref{tab:tableCPevenII}.   The branching fraction for $\tau \rightarrow \mu \rho$ is obtained by a cut on the invariant mass of the pair of charged pions ($\pi^+ \pi^-$), $ 587~\text{MeV} < \sqrt{s} <962~\text{MeV}$~\cite{Miyazaki:2011xe}.  The processes $\tau \rightarrow \mu \gamma$ and $\tau \rightarrow 3 \mu$ receive the dominant contribution from two-loop diagrams of the Barr-Zee type~\cite{Chang:1993kw}. We find our results for these processes to be in good agreement with those of Ref.~\cite{Harnik:2012pb}.  Even though the experimental limits for $\tau \rightarrow \mu \gamma$ and $\tau \rightarrow 3 \mu$ are very similar, the extracted bound from $\tau \rightarrow 3 \mu$ is weaker by an order of magnitude mainly due to the additional factor of $\alpha_{\text{em}}$.   

The $90 \%$~CL current upper bounds on $\mathrm{BR}( \tau \rightarrow \ell \pi^+ \pi^-)$ set by the Belle collaboration are at the $10^{-8}$ level using $854$~fb$^{-1}$ of collected data~\cite{Miyazaki:2012mx}.  
While  weaker by one order of magnitude compared to $\tau \to \ell \gamma$,  the bounds from 
$\tau \to \ell \pi \pi$ are quite less sensitive to the UV detail of the theory, and thus allow one to 
probe more directly the Higgs LFV couplings.  We observe that
the Belle and BaBar collaborations have not presented  results for the $\tau \rightarrow \ell \pi^0 \pi^0$ mode.  The current upper limit in this channel was set by the CLEO collaboration with $4.68$~fb$^{-1}$ of collected data, $\mathrm{BR}( \tau \rightarrow \mu \pi^0 \pi^0) < 1.4 \times 10^{-5} $ and  $\mathrm{BR}( \tau \rightarrow e \pi^0 \pi^0) < 6.5 \times 10^{-6} $ at $90 \%$ CL~\cite{Bonvicini:1997bw}. 

 Note from Fig.~\ref{taumupipitotal} that the Higgs and photon mediated contributions to $\tau \rightarrow \mu \pi^+ \pi^-$ are of similar size.  One can then infer that if the mode $\tau \rightarrow \mu \pi^0 \pi^0$ had been updated by the Belle or BaBar collaborations, it would be possible to set a limit on $\sqrt{ |Y^{h}_{\mu \tau}|^2 + |Y^{h}_{\tau \mu}|^2}$ at the $10^{-1}$ level from this process.  This is reinforced by the fact that the upper-limit set by the CLEO collaboration on the mode $\tau \rightarrow \mu \pi^+ \pi^-$ is very similar to that for $\tau \rightarrow \mu \pi^0 \pi^0$~\cite{Bonvicini:1997bw}.  The process $\tau \rightarrow \mu \pi^0 \pi^0$ has the advantage compared to $\tau \rightarrow \mu \pi^- \pi^+$, that it cannot be mediated by the photon and is therefore not affected by possible NP effects entering into the effective dipole operator.  The decay $\tau \rightarrow \mu \pi^0 \pi^0$ establishes the most direct connection between searches for LFV $\tau$ decays at B-factories and the search for LFV Higgs decays at the LHC: $pp (gg) \rightarrow h \rightarrow \tau \mu$.    Similar arguments apply for $\tau \rightarrow e \pi^0 \pi^0$.    If LFV Higgs decays are observed at some point, this would imply lower bounds on the  $\tau \rightarrow \ell \pi^0 \pi^0$ BR.   We therefore encourage the experimental collaborations to provide limits for this channel in the future.

\begin{table}[t!]\begin{center}
\vspace{0.2cm}
\begin{tabular}{|c|c|c|c|c|c|}
\hline
Process &  $(\mathrm{BR \times 10^{8}} )~90\%$ CL&   $ \rule{0cm}{0.5cm} \sqrt{ |Y^{h}_{\mu \tau}|^2 + |Y^{h}_{\tau \mu}|^2  }$  & Operator(s)  \\ 
\hline  \hline
$\tau \rightarrow \mu \gamma$ &  $<4.4$ ~\cite{Aubert:2009ag}  & $<0.016$    &  Dipole\\  \hline
$\tau \rightarrow \mu \mu \mu$ &  $<2.1$ ~\cite{Hayasaka:2010np}   & $<0.24$   & Dipole \\  \hline
$\tau \rightarrow \mu \pi^+ \pi^-$ &  $<2.1$ ~\cite{Miyazaki:2012mx}     & $<0.13$  & Scalar, Gluon, Dipole  \\  \hline
$\tau \rightarrow \mu \rho$ &  $< 1.2$ ~\cite{Miyazaki:2011xe}   &    $<0.13$  & Scalar, Gluon, Dipole  \\  \hline
$\tau \rightarrow \mu \pi^0 \pi^0$ &  $< 1.4 \times 10^{3}$ ~\cite{Bonvicini:1997bw} &    $<6.3$ & Scalar, Gluon   \\   
\hline
\end{tabular}
\caption{\it \small Current experimental upper bounds on the different processes considered as well as the bounds obtained on $\sqrt{ |Y^{h}_{\mu \tau}|^2 + |Y^{h}_{\tau \mu}|^2  }$ for a CP-even Higgs at $125$~GeV and SM-like diagonal couplings  $y_{f}^{h}=1$.   
The last column indicates the dominant operators contributing to each process. }
\label{tab:tableCPeven}
\end{center}\end{table}

\begin{table}[t!]\begin{center}
\vspace{0.2cm}
\begin{tabular}{|c|c|c|c|c|c|}
\hline
Process &  $(\mathrm{BR \times 10^{8}} )~90\%$ CL&   $ \rule{0cm}{0.5cm} \sqrt{ |Y^{h}_{e \tau}|^2 + |Y^{h}_{\tau e}|^2  }$ & Operator(s)  \\ 
\hline  \hline
$\tau \rightarrow e \gamma$ &  $<3.3$ ~\cite{Aubert:2009ag}  & $<0.014$  & Dipole  \\  \hline
$\tau \rightarrow e e e$ &  $<2.7$ ~\cite{Hayasaka:2010np}   & $<0.12$ & Dipole \\  \hline
$\tau \rightarrow e \pi^+ \pi^-$ & $ < 2.3$ ~\cite{Miyazaki:2012mx}  & $<0.14$  & Scalar, Gluon, Dipole \\ \hline
$\tau \rightarrow e \rho$ & $ < 1.8$ ~\cite{Miyazaki:2011xe} & $<0.16$  &Scalar, Gluon, Dipole  \\  \hline
$\tau \rightarrow e \pi^0 \pi^0$ &  $< 6.5 \times 10^{2}$ ~\cite{Bonvicini:1997bw} &    $<4.3$   & Scalar, Gluon  \\  
\hline
\end{tabular}
\caption{\it \small Current experimental upper bounds on the different processes considered as well as the bounds obtained on $\sqrt{ |Y^{h}_{e \tau}|^2 + |Y^{h}_{\tau e}|^2  }$ for a CP-even Higgs at $125$~GeV and SM-like diagonal couplings  $y_{f}^{h}=1$. 
The last column indicates the dominant operators contributing to each  process.  }
\label{tab:tableCPevenII}
\end{center}\end{table}

\subsubsection{The impact of hadronic matrix elements}   \label{sec:discu}
\begin{figure}[ht!]
\centering
\includegraphics[width=7cm,height=6.cm]{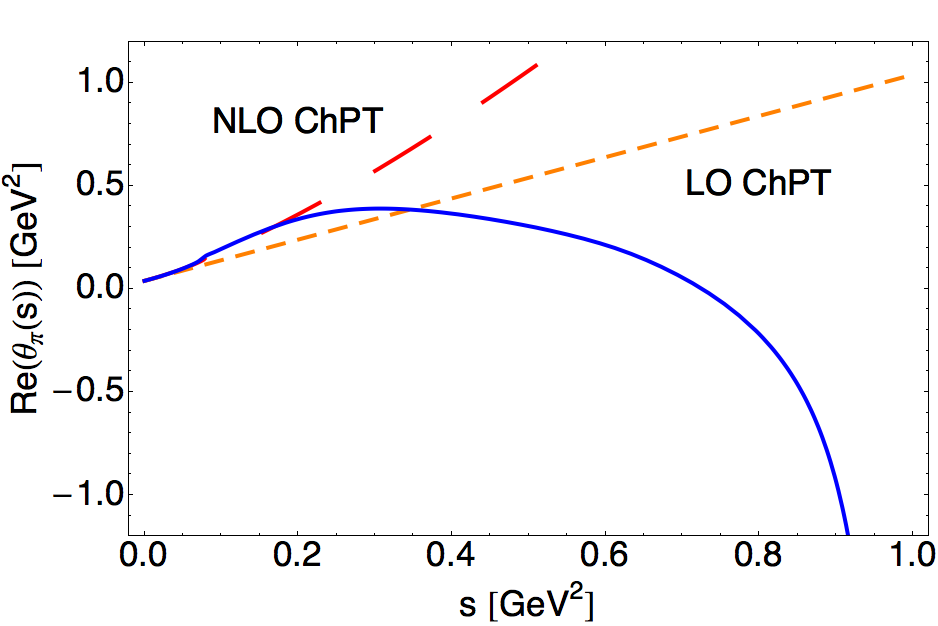}
~~~~
\includegraphics[width=7cm,height=6cm]{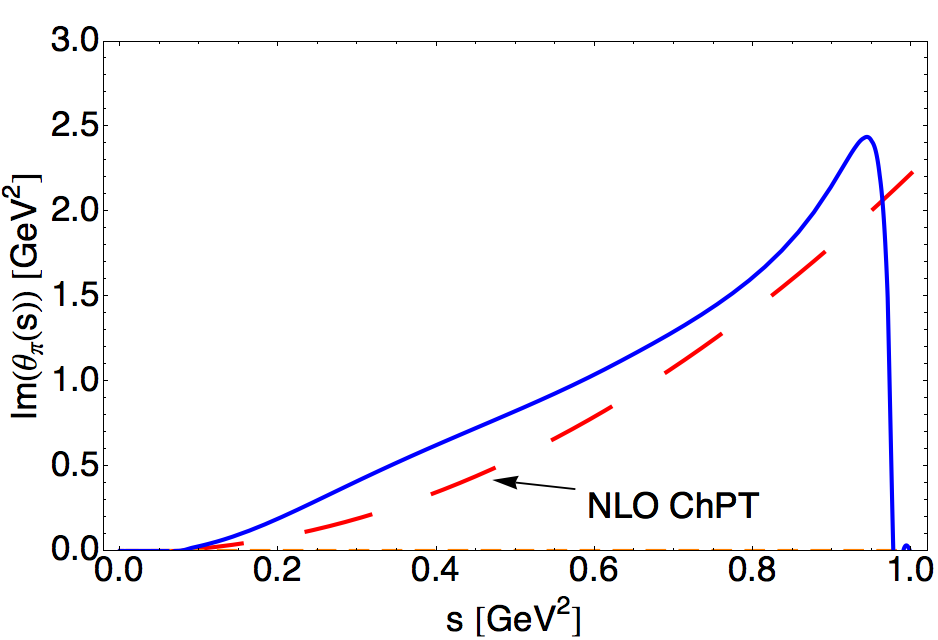}
\caption{\label{formfactors_fig} \it \small  Real (left) and imaginary (right) part of the form factor associated with the trace of the QCD energy-momentum tensor, $\theta_{\pi}(s)$, using different treatments: LO-ChPT (short-dashed orange), NLO-ChPT (long-dashed red), and our prediction based on ChPT and dispersion relations (continuous blue). }
\end{figure}

\begin{figure}[ht!]
\centering
\includegraphics[width=7cm,height=6.cm]{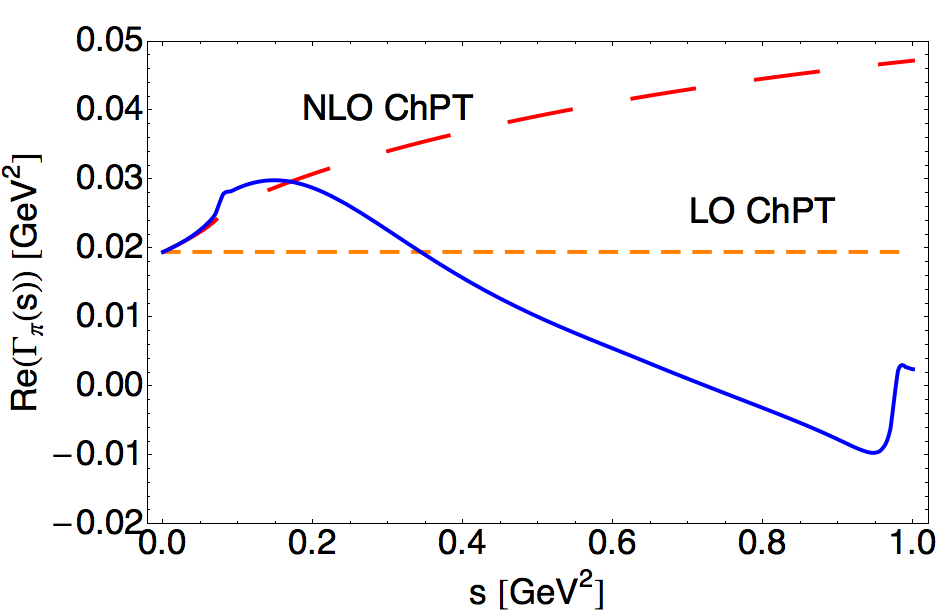}
~~~
\includegraphics[width=7cm,height=6cm]{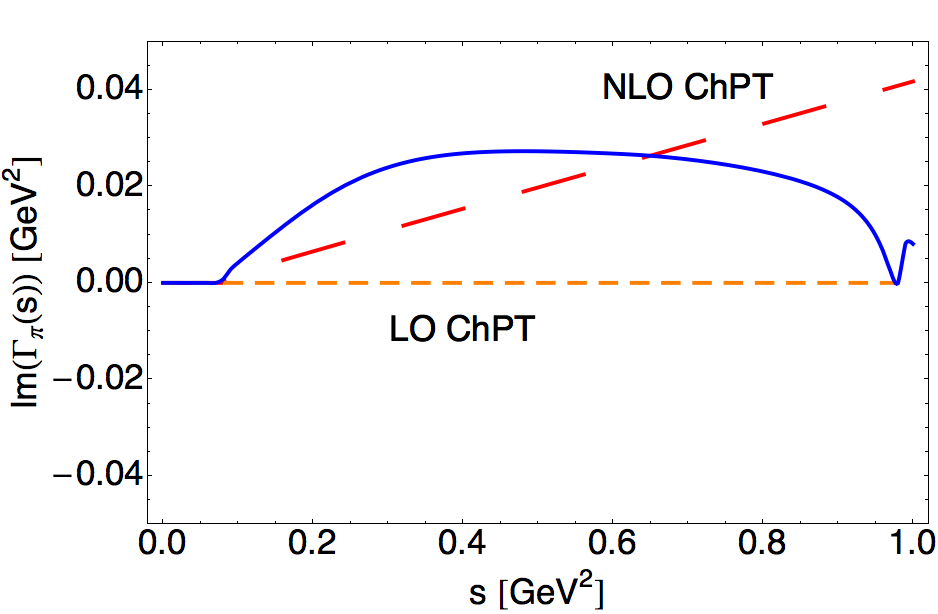}  \\
\includegraphics[width=7cm,height=6.cm]{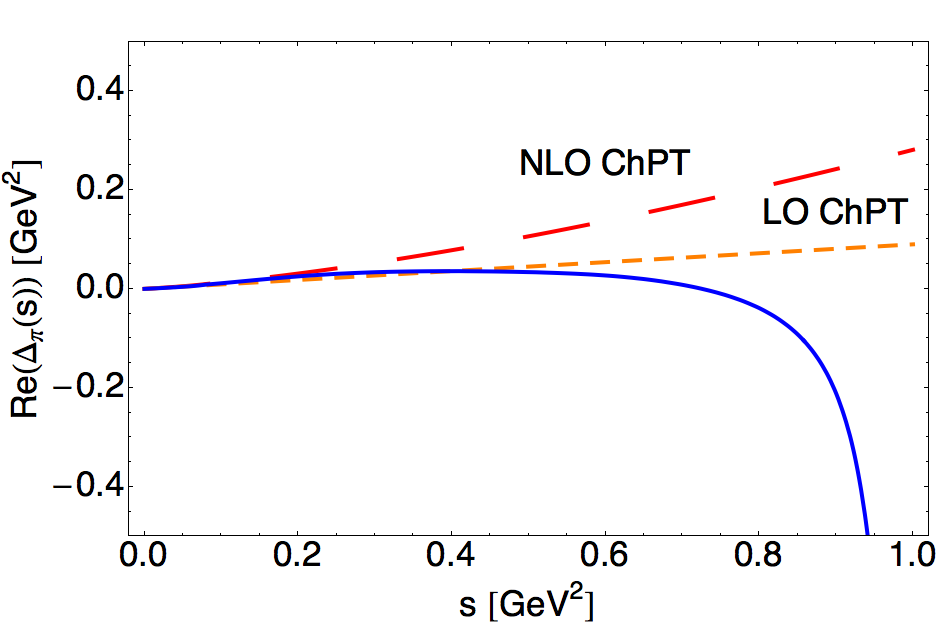}
~~~ \qquad
\includegraphics[width=7cm,height=6cm]{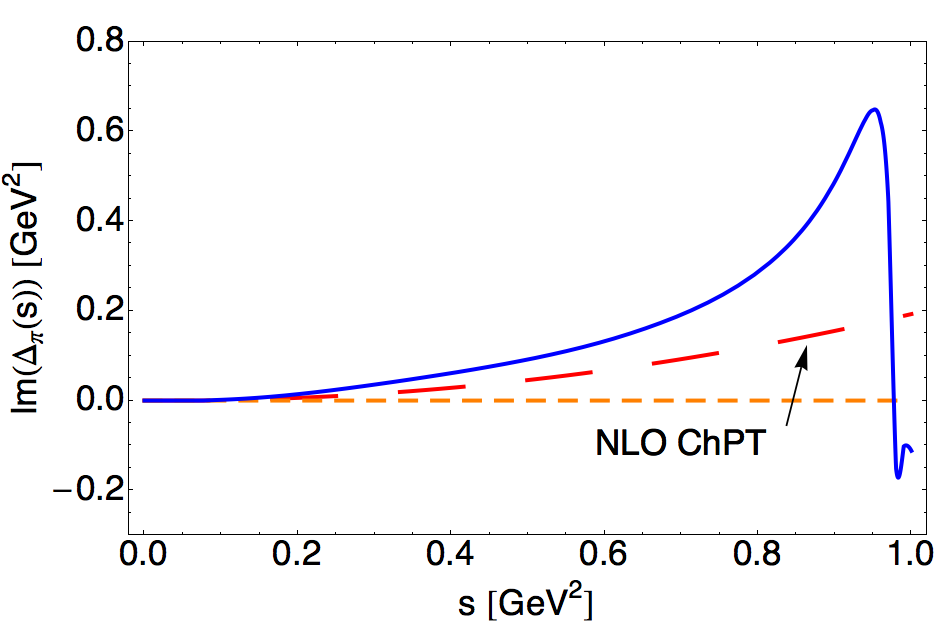}
\caption{\label{formfactors_figII} \it \small  Real (left) and imaginary (right) part for the scalar form factors $(\Gamma_{\pi}(s), \Delta_{\pi}(s))$, using different treatments: LO-ChPT (short-dashed orange), NLO-ChPT (long-dashed red), and  our prediction based on ChPT and dispersion relations (continuous blue). }
\end{figure}

In Fig.~\ref{taumupipitotal} (right-panel) we show the branching ratio for $\tau \rightarrow \mu \pi^+ \pi^-$ considering only the LO-ChPT prediction for the form factor $\Gamma_{\pi}(s)^{\text{\tiny{LO-ChPT}}} = m_{\pi}^2$ (while neglecting $\Delta_{\pi}(s)$ and $\theta_{\pi}(s)$) as done in Refs.~\cite{Black:2002wh,Kanemura:2005hr,Arganda:2008jj}.   Our prediction considering the three form factors $\{ \Gamma_{\pi}(s), \Delta_{\pi}(s), \theta_{\pi}(s) \}$, estimated using ChPT together with dispersion relations, is observed to be significantly larger.  It is important to clarify some points regarding such comparison between our results and those that have been considered previously by other authors using LO-ChPT.  First, a proper treatment of the decay $\tau \rightarrow \mu \pi^+ \pi^-$ would require taking into account not only $\Gamma_{\pi}(s)$ as is usually done, but also $\theta_{\pi}(s)$ and $\Delta_{\pi}(s)$ which actually provide the dominant contributions to the decay rate.   The LO-ChPT prediction for these form factors is~\cite{Donoghue:1990xh}
\be  \label{ffsLOChPt}
\theta_{\pi}(s)^{\text{\tiny{LO-ChPT}}} =  s + 2 m_{\pi}^2 \,, \qquad \Gamma_{\pi}(s)^{\text{\tiny{LO-ChPT}}} = m_{\pi}^2  \,, \qquad \Delta_{\pi}(s)^{\text{\tiny{LO-ChPT}}} =  d_F s + d_B m_{\pi}^2  \,. 
\ee
Here $d_F=0.09$ while $d_B \simeq 0$, and we refer the reader to  Ref.~\cite{Donoghue:1990xh} for the respective NLO-ChPT predictions.  The range of validity of the ChPT form factors is about $\sqrt{s_\chi} \sim 0.3$~GeV (LO-ChPT) or $\sqrt{s_\chi} \sim 0.5$~GeV (NLO-ChPT), see Figs.~\ref{formfactors_fig}-\ref{formfactors_figII}.  The LO-ChPT form factors are always real. The absorptive contribution starts at  
NLO in ChPT due to the appearance of re-scattering one-loop diagrams  generated by interaction terms of the leading chiral Lagrangian~\cite{Donoghue:1990xh}.    Above $\sqrt{s} \sim 0.5$~GeV, even the NLO-ChPT is not reliable anymore and significant departures can be observed compared with the form factors obtained using dispersion relations, 
that  take into account  large  $\pi\pi$  rescattering effects beyond  one-loop.

In order to asses the impact of the hadronic matrix elements for the calculation of the $\tau \rightarrow \ell \pi \pi$ decay rate, we consider the ratio
\begin{equation}\label{ratio_eq}
  R \;=\; \dfrac{   \int_{s_{\text{min}}}^{s_\chi} ds \,  \left[   d \Gamma(\tau \rightarrow \mu \pi^+ \pi^-)_{ \text{Higgs}}/ds \right]_{\text{LO-ChPT}}  }{ \int_{s_{\text{min}}}^{s_{\text{max}}} ds \, \left[d \Gamma(\tau \rightarrow \mu \pi^+ \pi^-)_{ \text{Higgs}}/ds \right]_{\text{ChPT + DR}} \,  }  \simeq 3.3 \times 10^{-5}  \,.
\end{equation}
Here  the numerator stands for the decay width calculated using the LO-ChPT predictions for the hadronic form factors (using the expressions in Eq.~\eqref{ffsLOChPt}), integrated up to a cut-off $\sqrt{s_\chi} \simeq 0.3$~GeV that specifies the range of validity of the LO-ChPT treatment. The denominator represents the decay width calculated using the form factors obtained in this work using dispersion relations (DR)  to extend the range of validity of the hadronic matrix elements to higher energies.       
The smallness of  $R \simeq 3.3 \times 10^{-5}$  shows the importance of  a proper treatment of the hadronic matrix elements. 
One may argue that by cutting the phase space integral at $\sqrt{s_\chi} = 0.3$~GeV one is throwing away most of the effect, some authors for example use the LO-ChPT estimates of the form factors over the full parameter space.  If we set $s_\chi = s_{\text{max}}$ in Eq.~\eqref{ratio_eq} we get instead the much larger value $R = 0.45$.  
This however is a very misleading result, 
based on using  LO-ChPT form factors in a kinematical regime where they no longer describe properly the hadronic dynamics.  
\footnote{
The authors of Ref.~\cite{Petrov:2013vka}, working within an effective theory framework,  
set bounds on  LFV  gluonic operators from $\tau \rightarrow \ell \pi \pi$ using the  LO-ChPT 
result in the chiral limit,  namely $\theta_\pi (s) = s$.
The claim that large departures from the LO-ChPT predictions are not to be expected~\cite{Petrov:2013vka} 
misses the fact that even in the chiral limit  ($m_u = m_d = m_s = 0$) ChPT is inadequate to 
describe the hadronic dynamics for large invariant masses of the $\pi\pi$ system $\sqrt{s} \sim 1$~GeV.
}

\subsection{A CP-odd Higgs with LFV couplings}
\label{sec:CPodd}
The Higgs boson at 125 GeV cannot be a pure pseudoscalar state, the experimental data already constrain its coupling to vector bosons to be very close to the SM value and analyses of the angular distributions in the Higgs decay final states also disfavor this possibility~\cite{CMS:yva,ATLAS:2013sla}.   Assuming that the $h(125)$ boson is the lightest CP-even state of a general 2HDM, current LHC and Tevatron measurements of the $h(125)$ properties imply that $g_{hVV} \simeq g_{hVV}^{\mathrm{SM}}$~\cite{Altmannshofer:2012ar,Celis:2013rcs,Barger:2013ofa,Lopez-Val:2013yba}.    Lepton flavor violating Yukawa couplings of $h$ would take the form: $(\Pi_{f})_{ij}/v \sin \tilde \alpha$ (see Eq.~\eqref{p1}) and are therefore suppressed by the small factor $ |\sin \tilde \alpha| = \left(1 - (g_{hVV}/ g_{hVV}^{\mathrm{SM}})^2 \right)^{1/2} $.   The LFV Yukawa couplings of the heavier CP-even state $H$ and the CP-odd Higgs $A$ on the other hand do not receive this suppression.  It is therefore interesting not only to consider searches for LFV decays of the $125$ GeV boson at the LHC, but also of possible additional Higgs bosons.    The question of which observables measurable at flavor factories could be related to the process $pp(gg) \rightarrow A \rightarrow \tau \mu$ relevant at the LHC then arises naturally. We argue in this section that the semileptonic 
decays  of $\tau$ into a pseudoscalar meson $P$,  $\tau \rightarrow \ell P$, provide a direct connection with the search for a CP-odd Higgs with LFV couplings at the LHC. 
In this section we focus on the CP-odd boson $A$, which implies a somewhat different  phenomenology of $\tau$ LFV decays compared
to the  CP-even state already analyzed in the previous section and in Refs.~\cite{Blankenburg:2012ex,Harnik:2012pb,Davidson:2012ds}.    
We do not consider the effect of interfering contributions of the different scalars $\varphi_k = \{h, H, A \}$ in $\tau \rightarrow \ell \gamma$ or the phenomenology of the charged Higgs, these have been discussed elsewhere~\cite{Chang:1993kw,Kanemura:2005hr,Davidson:2010xv}.

At the relevant energy scale for $\tau$ decays, the heavy-quarks can be integrated out from the theory, the effective Lagrangian describing the interactions of the CP-odd Higgs with the light quarks is then given by~\cite{Shifman:1978zn}
\begin{equation} \label{effLagII}
\mathcal L_{eff}^{A}  \simeq - \frac{A}{v}  \, \left(  \, \sum_{q=u,d,s}\, y_{q}^{A} \, m_{q} \, \bar q  i  \gamma_{5}  \,q   -  \sum_{q=c,b,t}\,  
y_{q}^{A} \,  
  \dfrac{\alpha_s}{8 \pi} \,  \,G_{\mu \nu}^{a} \, \widetilde{G}_{\mu \nu}^{a} \right)   \,,
\end{equation}
with real couplings $y_q^A$ related to those  of Eq.~(\ref{laggen})   by ${\rm Im} Y_{qq}^A = (m_q/v) y_q^A$
and the dual tensor of the gluon field strength defined by  $\widetilde G^{a}_{\rho \nu} = \frac{1}{2}\, \epsilon_{\rho \nu \alpha \beta}  \,G^{a}_{\alpha \beta}$ (with $\epsilon^{0123}=1$).    Contrary to a CP-even Higgs boson, a CP-odd Higgs with LFV couplings can mediate at tree-level the semileptonic decays $\tau \rightarrow \ell P$, where $P= \pi^0, \eta, \eta^{\prime}$, stands for a pseudoscalar meson.  Semileptonic $\tau$ decays into a pseudoscalar meson cannot be mediated by the photon either, so that this mode isolates the CP-odd Higgs exchange.  The decays $\tau \rightarrow \ell P$ therefore establish a direct connection with the search for CP-odd Higgs decays into LFV channels at the LHC, with the Higgs being produced via gluon fusion. The relevant hadronic matrix elements can be obtained following the FKS mixing scheme~\cite{Feldmann:1998vh,Beneke:2002jn}, those involving the Higgs coupling to light-quarks are parametrized by  
\begin{align}  \label{mesonII}
 \langle \pi^0(p)  |  \bar u  \, \gamma_5 \, u   |  0 \rangle &\;=\;    i \, \frac{m_{\pi}^2}{2 \sqrt{2} \hat m} \, f_{\pi}  \,,  \qquad  \qquad  \langle \pi^0  |  \bar d \, \gamma_5  \,d   | 0 \rangle = - \langle \pi^0(p)  |  \bar u  \, \gamma_5 \, u   |  0 \rangle  \,,   \\ \nonumber 
  \langle \eta^{(\prime)}(p)   |   \bar q \, \gamma_5 \, q |  0 \rangle &\;= \;  - \frac{i}{2 \sqrt{2} m_q   }\, h_{\eta^{(\prime)} }^{q}    \,, \qquad   \, \, \,
  \langle \eta^{(\prime)} (p) |   \bar s \, \gamma_5 \, s |  0 \rangleÊ \;= \;  - \frac{i}{2 m_s   }\, h_{\eta^{(\prime)}}^{s}    \,,   
 \end{align}
while those related to the loop-induced effective operator $A G_{\mu \nu}^{a} \, \widetilde{G}_{\mu \nu}^{a} $ are given by
\begin{align}  \label{omega}
 \langle  \eta^{(\prime)}(p)  |  \frac{ \alpha_s}{4 \pi}  \, G_a^{\mu \nu}  \widetilde{G}_{\mu \nu}^{a}   | 0  \rangle &=    a_{\eta^{(\prime)}}    \,.
  \end{align}
Numerical values for the different parameters appearing in Eqs.~(\ref{mesonII},\ref{omega}) are given in Tab.~\ref{tab:inputs}.  The contributions from the effective operator $A G_{\mu \nu}^{a} \, \widetilde{G}_{\mu \nu}^{a} $ to the decay $\tau \rightarrow \ell \pi$ vanishes in the isospin limit $m_u= m_d$~\cite{Gross:1979ur} and we do not consider it here.   
\begin{table}[ht!]\begin{center}
\vspace{0.2cm}
\begin{tabular}{|c|c| }
\hline 	
Parameter & Value \\
\hline  \hline
$f_{\pi}$  &   $130.41  \pm 0.20$~MeV \\
$h_{\eta}^{q}$ &  $ 0.001 \pm 0.003$~GeV$^3$   \\
$h_{\eta^{\prime}}^{q}$ & $ 0.001 \pm 0.002$~GeV$^3$   \\
$h_{\eta}^{s}$ &  $ -0.055 \pm 0.003$~GeV$^3$  \\
$h_{\eta^{\prime}}^{s}$ & $ 0.068 \pm 0.005$~GeV$^3$   \\
$a_{\eta}$ &   $0.022 \pm 0.002$~GeV$^3$   \\
$a_{\eta^{\prime}}$ &  $ 0.056 \pm 0.002$~GeV$^3$ \\
\hline
\end{tabular}
\caption{\it \small Numerical values for the hadronic matrix elements relevant for $\tau \rightarrow \ell P$ ($P = \pi, \eta, \eta^{\prime}$) obtained in the FKS mixing scheme~\cite{Feldmann:1998vh,Beneke:2002jn}.}
\label{tab:inputs}
\end{center}\end{table}
The total decay width for $\tau \rightarrow \ell \pi^0$, neglecting small lepton and pion mass effects, reads
\be
\Gamma(\tau \rightarrow \ell \pi^0) \;=\;  \frac{ f_{\pi}^2 \, m_{\pi}^4 \, m_{\tau}  }{256 \pi M_A^4 \, v^2} \left( |Y^{A}_{\tau \mu}|^2 + |Y^{A}_{\mu \tau}|^2 \right)  
 \left( y_u^{A} - y_{d}^{A}  \right)^2   \,,
\ee
the amplitude for $\tau \rightarrow \mu \pi^0$ vanishes exactly in the limit $y_u^{A} = y_d^{A}$ since the $\pi^0$ only selects the isovector component of the amplitude.  
The decay width for $\tau \rightarrow \ell \eta$ can be written using the definitions of Eqs.~(\ref{mesonII},\ref{omega}) as (neglecting small lepton mass effects)
\be \label{taumueta}
\Gamma(\tau \rightarrow \ell \eta) =  
\frac{
 \bar{\beta} 
\, (m_{\tau}^2 - m_{\eta}^2)  \left(   |  Y_{\mu \tau}^{A} |^2 + | Y_{\tau \mu}^{A} |^2   \right)  }{ 256 \,\pi \,M_A^4\, v^2\, m_{\tau}}        
\Bigl[  (y_u^A + y_d^A)  h_\eta^q    + \sqrt{2}  y_s^A   h_\eta^s   - \sqrt{2}  a_\eta    \sum_{q=c,b,t}   \, y_{q}^{A}   \Bigr]^2  
~, 
\ee
with $\bar{\beta} = (1 - 2 (m_\ell^2 + m_\eta^2)/m_\tau^2    +  (m_\ell^2 - m_\eta^2)^2/m_\tau^4 )^{1/2}$.
A simple replacement of $\eta \rightarrow \eta^{\prime}$ in Eq.~\eqref{taumueta} gives the expression for $\Gamma( \tau \rightarrow \ell \eta^{\prime})$. 

  A CP-odd Higgs boson would also give rise to an effective dipole operator at the loop-level~\cite{Chang:1993kw}, contributing then to $\tau \rightarrow \ell \gamma$, and photon-mediated $\tau \rightarrow \ell \pi^+ \pi^-, \, 3 \ell$ decays.  Note that while $\tau\rightarrow 3 \ell$ is also mediated at tree-level by the CP-odd Higgs, the semileptonic decay $\tau \rightarrow \ell \pi \pi$ are not due to spin-parity conservation.  The CP-odd Higgs exchange contribution to $\tau \rightarrow 3 \mu$ is however sub-leading compared to that coming from the two-loop diagrams of the Barr-Zee type due to the small Yukawa coupling to the muons, see the recent discussion in Ref.~\cite{Harnik:2012pb}.  In Tables~\ref{tab:tableCPodd} and \ref{tab:tableCPoddII}  we summarize the bounds on $\sqrt{ |Y^{A}_{\ell \tau}|^2 + |Y^{A}_{\tau \ell}|^2  }$ from the different $\tau$ decays considered fixing the diagonal couplings to $|y_{f}^{A}|=1$.   The scaling of the semileptonic $\tau \rightarrow \ell P$  decay rates with the CP-odd Higgs mass is very simple, $\propto M_A^{-4}$, while that for processes mediated by the photon is non-trivial due to loop-functions entering in the calculation of the transition dipole moment. The stringent bound coming from $\tau \rightarrow \ell \gamma$ is sensitive to possible interference effects from other scalars or heavy particles from a UV completion of the theory.   The semileptonic decays $\tau \rightarrow \ell P$ on the other hand are mediated at tree-level by the CP-odd Higgs exchange and provide then a more reliable bound in this respect.

 In case any LFV signal $pp \rightarrow \tau \ell + X$ is observed at the LHC, it will be crucial to determine the properties of the mediator.    The complementarity between low-energy searches for LFV $\tau$ decays and the energy frontier is very important for this purpose.  A CP-even Higgs with LFV couplings for example would give rise to $\tau \rightarrow \ell \gamma$ decays via loop contributions while it cannot mediate semileptonic $\tau \rightarrow \ell P$ decays.    If the 125 GeV Higgs turns out to have sizable LFV couplings and $h \rightarrow \tau \ell$ decays are observed at some point at the LHC, specific patterns between all the possible LFV $\tau$ decays would then be predicted and any departure from these could be an indication of additional particles with LFV couplings for example.

\begin{table}[ht!]\begin{center}
\vspace{0.2cm}
\begin{tabular}{|c|c|c|c|c|c|}
\hline 
Process &  $(\mathrm{BR \times 10^{8}} )~90\%$ CL&   $M_A = 200$~GeV  & $M_A = 500$~GeV   &  $M_A = 700$~GeV   \\ 
\hline  \hline
$\tau \rightarrow \mu \gamma$ &  $<4.4$ ~\cite{Aubert:2009ag} &  $Z <  0.018$  &  $Z < 0.040$   &  $Z <  0.055$   \\ \hline
$\tau \rightarrow \mu \mu \mu$ &  $<2.1$ ~\cite{Hayasaka:2010np}    &  $Z <  0.28$  & $Z <  0.60$  & $Z <  0.85$ \\ \hline
$^{(*)}$  $\tau \rightarrow \mu \pi$  &  $<11$ ~\cite{Aubert:2006cz}   &   $Z <  41$  & $Z <  257$  & $Z <  503$ \\  \hline
$^{(*)}$  $\tau \rightarrow \mu \eta$  &  $<6.5$ ~\cite{Aubert:2006cz}  &   $Z <  0.52$  & $Z <  3.3$  & $Z <  6.4$ \\  \hline
$^{(*)}$  $\tau \rightarrow \mu \eta^{\prime}$  &  $<13$ ~\cite{Aubert:2006cz}  &  $Z <  1.1$ &  $Z <  7.2$   &  $Z <  14.1$  \\ \hline
$\tau \rightarrow \mu \pi^+ \pi^-$ &  $<2.1$ ~\cite{Miyazaki:2012mx}   & $ Z<0.25$ & $ Z<0.54$ & $Z<0.75$  \\  \hline
$\tau \rightarrow \mu \rho$ &  $< 1.2$  ~\cite{Miyazaki:2011xe}  &  $Z<0.20$&   $Z<0.44$ & $Z<0.62$ \\   
\hline
\end{tabular}
\caption{\it \small Current experimental upper bounds on the different processes considered as well as the bounds obtained on $Z \equiv \sqrt{ |Y^{A}_{\mu \tau}|^2 + |Y^{A}_{\tau \mu}|^2  }$ for different values of the CP-odd Higgs mass and SM-like diagonal couplings  $|y_{f}^{A}|=1$. Neither the effective dipole operator nor the CP-even Higgs exchange contribute to the processes marked with $^{(*)}$.  }
\label{tab:tableCPodd}
\end{center}\end{table}

\begin{table}[ht!]\begin{center}
\vspace{0.2cm}
\begin{tabular}{|c|c|c|c|c|}
\hline 
Process &  $(\mathrm{BR \times 10^{8}} )~90\%$ CL&   $M_A = 200$~GeV  & $M_A = 500$~GeV   &  $M_A = 700$~GeV   \\ 
\hline  \hline
$\tau \rightarrow e \gamma$ &  $< 3.3$ ~\cite{Aubert:2009ag}    &  $Z <  0.016$  &  $Z < 0.034$   &  $Z <  0.05$   \\ \hline
$\tau \rightarrow e  e e$  &  $<2.7$ ~\cite{Hayasaka:2010np}     &  $Z <  0.14$  & $Z <  0.30$  & $Z <  0.42$ \\ \hline
$^{(*)}$  $\tau \rightarrow e \pi$ &  $< 8$ ~\cite{Aubert:2006cz}   &   $Z <  35$  & $Z <  219$  & $Z <  430$ \\  \hline
$^{(*)}$  $\tau \rightarrow e \eta$ &  $< 9.2$ ~\cite{Aubert:2006cz}   &   $Z <  0.6$  & $Z <  3.9$  & $Z <  7.6$ \\  \hline
$^{(*)}$  $\tau \rightarrow e \eta^{\prime}$&  $<16$ ~\cite{Aubert:2006cz}   &  $Z <  1.3$ &  $Z <  8$   &  $Z <  15.6$  \\ \hline
$\tau \rightarrow e \pi^+ \pi^-$ &  $<2.3$ ~\cite{Miyazaki:2012mx}   & $ Z<0.26$ & $ Z<0.56$ & $Z<0.80$  \\  \hline
$\tau \rightarrow e \rho$ &  $< 1.8$ ~\cite{Miyazaki:2011xe}    &  $Z<0.25$&   $Z<0.54$ & $Z<0.76$ \\   
\hline
\end{tabular}
\caption{\it \small Current experimental upper bounds on the different processes considered as well as the bounds obtained on $Z \equiv \sqrt{ |Y^{A}_{e \tau}|^2 + |Y^{A}_{\tau e}|^2  }$ for different values of the CP-odd Higgs mass and SM-like diagonal couplings  $|y_{f}^{A}|=1$.  Neither the effective dipole operator nor the CP-even Higgs exchange contribute to the processes marked with $^{(*)}$. }  
\label{tab:tableCPoddII}
\end{center}\end{table}

\begin{figure}[ht!]
\centering
\includegraphics[width=7.5cm,height=5.5cm]{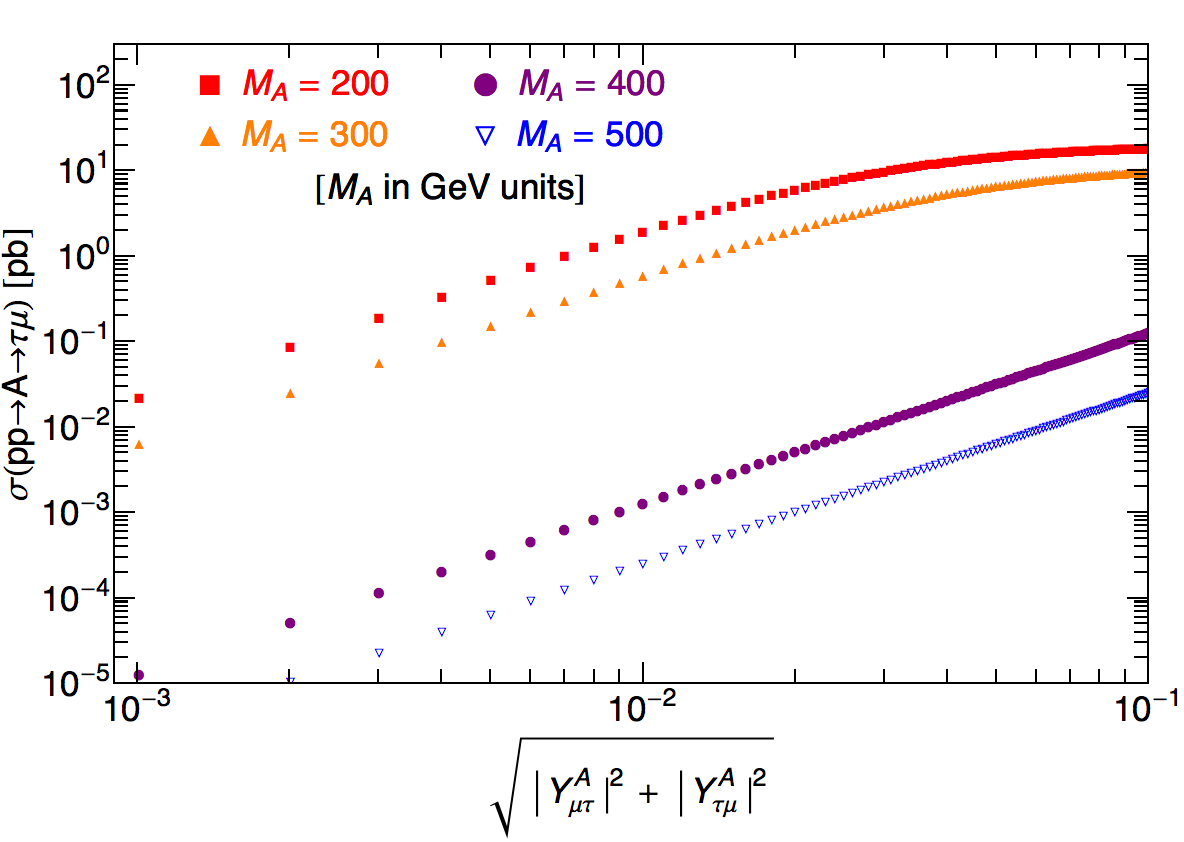}
~
\includegraphics[width=7.5cm,height=5.5cm]{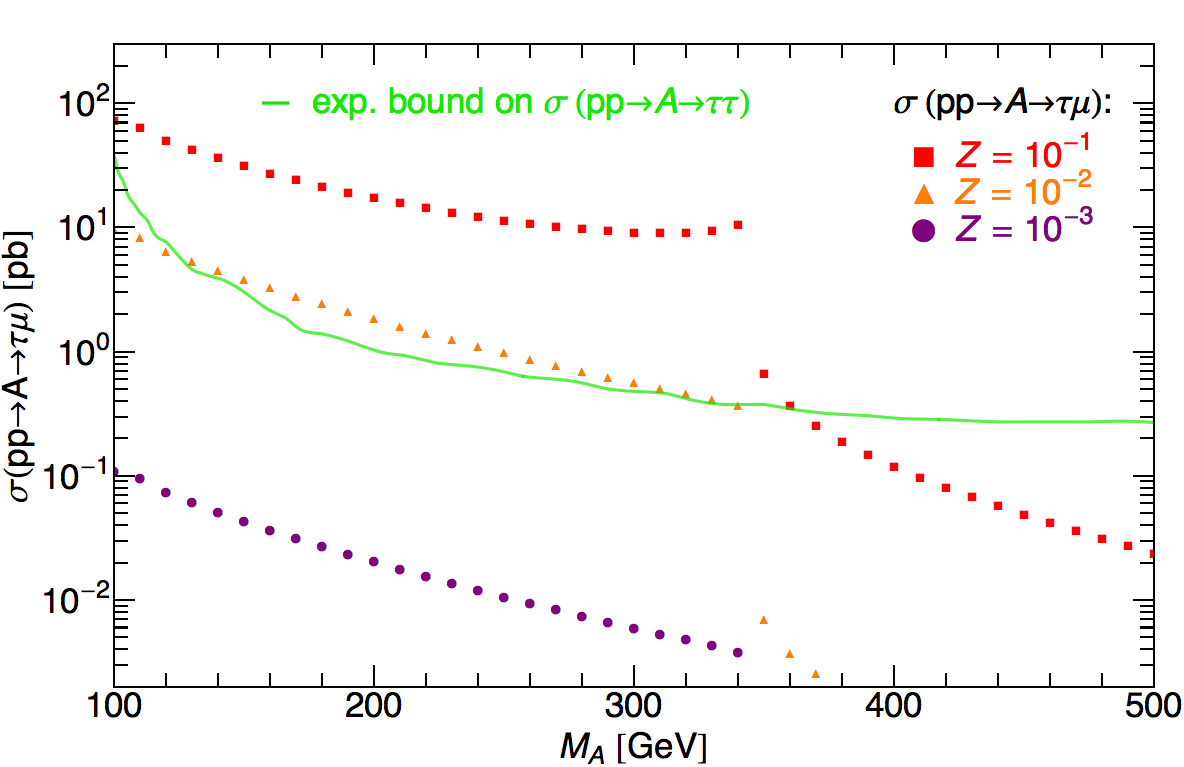}
\caption{\label{CPoddLHC} \it \small  Left-plot: Inclusive cross-section $\sigma(pp \rightarrow A \rightarrow \tau \mu)$ at $\sqrt{s}=8$~TeV as a function of $Z \equiv \sqrt{ |Y^{A}_{\mu \tau}|^2 + |Y^{A}_{\tau \mu}|^2  }$ for SM-like diagonal couplings $|y_{f}^{A}|=1$, taking different values of the CP-odd Higgs mass $M_A$.  Right-plot:  Inclusive cross-section $\sigma(pp \rightarrow A \rightarrow \tau \mu)$ at $\sqrt{s}=8$~TeV as a function of the CP-odd Higgs mass $M_A$ for  $  Z = 0.1$~(squares), $0.01$~(triangles), $0.001$~(circles).   The continuous green line shows the upper bound at $95\%$~CL on the cross-section $\sigma(pp \rightarrow A \rightarrow \tau \tau)$ using 4.7 $fb^{-1}$ of data by the ATLAS collaboration~\cite{Aad:2012yfa}.  }
\end{figure}

In view of the possibility to make a dedicated search for heavy scalars decaying to a $\tau-\mu$ pair, similar to the flavor conserving searches~\cite{Aad:2012yfa}, we estimate the total cross-section for $\sigma(pp \rightarrow A \rightarrow \tau \mu)$ at the LHC.   The inclusive Higgs production cross-section $\sigma(pp \rightarrow A)$ was obtained using  the \textit{SusHi} code \cite{Harlander:2012pb,Harlander:2002wh}, considering only the dominant gluon fusion production mode. Higgs partial decay widths were obtained using the \textit{2HDMC} code~\cite{Eriksson:2009ws}.  For a CP-odd Higgs we have (neglecting small lepton mass corrections)
\be
\Gamma(A \rightarrow \tau^+ \mu^- + \tau^- \mu^+) \equiv \Gamma(A \rightarrow \tau \mu ) = \frac{   M_A \left(  |Y^{A}_{\tau \mu}|^2 + |Y^{A}_{\mu \tau}|^2   \right)    }{ 8 \pi } \,.
\ee
We assume that besides $ A \rightarrow \tau \mu$, only SM decay channels are significant ($A \rightarrow gg, \bar c c, \bar b b, \tau \tau, \ldots$) and we fix the diagonal Yukawa couplings at  $|y_{f}^{A}| =1$.    Large branching ratios for the fermionic decays of a CP-odd Higgs and in particular for the $A \rightarrow \tau \mu$ mode, can be obtained since the CP-odd Higgs does not couples to $VV= W^+W^-, ZZ$ at tree-level.  Here we focus on the $\tau - \mu$ mode but an analogous analysis can be carried for $\tau -e$.

In Fig.~\ref{CPoddLHC} we show the total cross-section $\sigma(pp \rightarrow A \rightarrow \tau \mu)$ as a function of $\sqrt{ |Y^{A}_{\mu \tau}|^2 + |Y^{A}_{\tau \mu}|^2  }$ for $\sqrt{s}=8$~TeV.   A large drop in the total cross section can be observed when $M_A \gtrsim 2 m_t$ since $A \rightarrow \bar t t$ decays become kinematically open and suppress the branching ratio $\mathrm{BR}(A \rightarrow \tau \mu)$.   The total cross-section $\sigma(pp \rightarrow A \rightarrow \tau \mu)$ can be as large as $\sim1$ pb for a CP-odd Higgs with $M_A \sim  200$~GeV and $\sqrt{ |Y^{A}_{\mu \tau}|^2 + |Y^{A}_{\tau \mu}|^2  } \sim 10^{-2}$, which is allowed in principle by low-energy constraints, see Tab.~\ref{tab:tableCPodd}.  Current upper bounds for the flavor conserving cross-section $\sigma(pp \rightarrow A \rightarrow \tau \tau)$ by the ATLAS collaboration using 4.7 $fb^{-1}$ of data are also shown in Fig.~\eqref{CPoddLHC}.  The bound on $\sigma(pp \rightarrow A \rightarrow \tau \tau)$ is already at the $\sim 1$ pb level.   One can therefore expect that the sensitivity of the LHC to a plausible heavy Higgs with LFV couplings should be very good compared to flavor constraints, as previous analyses have shown for the 125 GeV boson~\cite{Harnik:2012pb,Davidson:2012ds}.   A detailed analysis of the LHC prospects to detect LFV Higgs decays of a heavy Higgs boson within the generic 2HDM was performed in Ref.~\cite{Assamagan:2002kf} finding promising results.

\section{Conclusions}
\label{sum}
The discovery of a new boson around $125$~GeV, $h(125)$, opens a new era in our understanding of the electroweak symmetry breaking mechanism, yet to be explored in detail.  Any departure from the SM Higgs properties or exotic effect associated with the $h(125)$ boson would be an indication of new physics beyond the SM.  The search for 
LFV phenomena in the scalar sector at the LHC has a special role in this respect, given 
the relatively weak constraints from low-energy experiments. 

While $h \rightarrow e \mu$ transitions are strongly constrained already by $\mu \rightarrow e \gamma$ decays and $\mu - e$ conversion in nuclei, the situation is completely different for $h \rightarrow \tau \ell$  ($\ell=e,\mu$) in which large decay rates are still allowed~\cite{Blankenburg:2012ex,Harnik:2012pb,Davidson:2012ds}.   The strongest bound on such LFV Higgs couplings is currently extracted from the radiative $\tau \rightarrow \ell \gamma$ decays.   This decay receives dominant contributions from two-loop diagrams of the Barr-Zee type due to the strong chirality suppression of the one-loop diagrams, making the bounds very sensitive to the underlying UV model.

Hadronic $\tau$-lepton decays offer an interesting low-energy handle to constrain possible LFV effects associated with the Higgs sector and in particular the $h(125)$ boson.   The bounds extracted from hadronic $\tau$-decays are less sensitive to the UV completion of the theory and establish a more direct connection with the search for LFV Higgs decays at the LHC.     We have shown in this work that the bounds obtained from semileptonic decays $\tau \rightarrow \ell \pi \pi$ are at the same level than those from $\tau \rightarrow 3 \mu$ ($\tau \rightarrow 3 \mu$ decays are dominated by the same two-loop diagrams than $\tau \rightarrow \ell \gamma$ and are therefore very sensitive to the UV completion of the theory).     This result was achieved thanks to an adequate description of the hadronic matrix elements involved, improving considerably over previous related works on this subject.     Concerning the semileptonic $\tau \rightarrow \ell \pi \pi$ transitions we emphasize the following results found in this work:

\begin{itemize}

\item 
In Sect.~\ref{sec::ffs} 
we provide a dispersive treatment of all the hadronic matrix elements needed to describe Higgs mediated $\tau \rightarrow \ell \pi \pi$ decays. 
These results  will be useful in analyzing $\tau \to \ell \pi \pi$ decays 
beyond the specific framework adopted here.

\item The form factors obtained in Sect.~\ref{sec::ffs} were used to extract robust bounds on possible LFV couplings of the $h(125)$ boson from current experimental data.  This was done in Sect.~\ref{sec:CPeven}, the main results being summarized in Tables~\ref{tab:tableCPeven} and \ref{tab:tableCPevenII}, 
as well as Figures~\ref{fig:spectra} and  \ref{taumupipitotal}.

\item We find that the dominant contributions to the Higgs mediated decay rate $\tau \rightarrow \ell \pi \pi$ arise from the effective Higgs couplings to gluons (induced by heavy quarks) and the strange component of the scalar current (Higgs coupling directly to strange quarks).  Previous treatments~\cite{Black:2002wh,Kanemura:2005hr,Arganda:2008jj} of these decays considering only the scalar current $m_u \bar u u + m_d \bar d d$ therefore do not capture the main contributions to the decay rate.

\item LO-ChPT predictions for the hadronic matrix elements contributing to $\tau \rightarrow \ell \pi \pi$ 
are valid only at very low energies $\sqrt{s} \lesssim 0.3$~GeV (see  Sect.~\ref{sec:discu}): if used over the whole phase space they lead to 
unreliable bounds on the LFV couplings.

\item 
Contrary to  $\tau \rightarrow  \ell  \pi^+ \pi^-$, the  $\tau \rightarrow  \ell  \pi^0 \pi^0$ 
decays cannot be mediated by  the effective dipole operator $(\bar \ell \sigma^{\mu \nu} P_{L,R} \tau) F_{\mu \nu}$ and isolate the CP-even 
Higgs exchange contribution.  
So  $\tau \rightarrow \ell  \pi^0 \pi^0$ decays  establish    the most direct connection between searches for LFV $\tau$ decays at B-factories and the search for LFV Higgs 
decays at the LHC ($pp (gg) \rightarrow h \rightarrow \tau \ell$). We  encourage the experimental collaborations to provide limits for  these  modes  in the future.

\end{itemize}

Finally, we point out in Sect.~\ref{sec:CPodd} that the search for LFV effects associated to the scalar sector should not be restricted to the $h(125)$ boson.  Within the general 2HDM, it is plausible that the LFV couplings of the $h(125)$ boson are too suppressed to be observed given that its coupling to vector bosons $VV=W^+W^-, ZZ$ is already constrained to be very close to the SM value.      If such an LFV extended scalar sector is realized in nature, it might be possible on the other hand to detect LFV phenomena at the LHC due to the decays of additional scalars for which such strong suppression of the LFV couplings does not takes place.   Current constraints from low-energy flavor experiments still allow for sizable LFV effects in this respect.

\subsection*{Acknowledgments}
We are very grateful to B.~Moussallam for 
illuminating discussions and for providing us with the solution of the Roy-Steiner equations needed as inputs to determine the scalar form factors.  We also thank 
Gilberto Colangelo, 
Heinrich Leutwyler, 
Antonio Pich,  and 
Jorge Portoles for useful discussions and comments.  A.C. would like to thank LANL Theoretical Division for its generous hospitality while part of this work was being done.    The work of A.C. is supported by the Spanish Ministry MECD through the FPU grant AP2010-0308.
The work of V.C and E.P. is supported by the DOE Office of Science and the LDRD program at Los Alamos National Laboratory.

\bibliographystyle{h-physrev}
\bibliography{HLFV}

\begin{thebibliography}{100}

\bibitem{CMS:yva}
CMS-Collaboration, 2013,
\newblock CMS-PAS-HIG-13-005.

\bibitem{ATLAS:2013sla}
ATLAS-Collaboration,
\newblock (2013),
\newblock ATLAS-CONF-2013-034, ATLAS-COM-CONF-2013-035.

\bibitem{Cheung:2013kla}
K.~Cheung, J.~S. Lee, and P.-Y. Tseng,
\newblock JHEP {\bf 1305}, 134 (2013), 1302.3794.

\bibitem{Ellis:2013lra}
J.~Ellis and T.~You,
\newblock JHEP {\bf 1306}, 103 (2013), 1303.3879.

\bibitem{Falkowski:2013dza}
A.~Falkowski, F.~Riva, and A.~Urbano,
\newblock (2013), 1303.1812.

\bibitem{Giardino:2013bma}
P.~P. Giardino, K.~Kannike, I.~Masina, M.~Raidal, and A.~Strumia,
\newblock (2013), 1303.3570.

\bibitem{Blankenburg:2012ex}
G.~Blankenburg, J.~Ellis, and G.~Isidori,
\newblock Phys.Lett. {\bf B712}, 386 (2012), 1202.5704.

\bibitem{DiazCruz:1999xe}
J.~L. Diaz-Cruz and J.~Toscano,
\newblock Phys.Rev. {\bf D62}, 116005 (2000), hep-ph/9910233.

\bibitem{Black:2002wh}
D.~Black, T.~Han, H.-J. He, and M.~Sher,
\newblock Phys.Rev. {\bf D66}, 053002 (2002), hep-ph/0206056.

\bibitem{Babu:2002et}
K.~Babu and C.~Kolda,
\newblock Phys.Rev.Lett. {\bf 89}, 241802 (2002), hep-ph/0206310.

\bibitem{Brignole:2004ah}
A.~Brignole and A.~Rossi,
\newblock Nucl.Phys. {\bf B701}, 3 (2004), hep-ph/0404211.

\bibitem{Kanemura:2005hr}
S.~Kanemura, T.~Ota, and K.~Tsumura,
\newblock Phys.Rev. {\bf D73}, 016006 (2006), hep-ph/0505191.

\bibitem{Paradisi:2005tk}
P.~Paradisi,
\newblock JHEP {\bf 0602}, 050 (2006), hep-ph/0508054.

\bibitem{Arganda:2008jj}
E.~Arganda, M.~Herrero, and J.~Portoles,
\newblock JHEP {\bf 0806}, 079 (2008), 0803.2039.

\bibitem{Herrero:2009tm}
M.~Herrero, J.~Portoles, and A.~Rodriguez-Sanchez,
\newblock Phys.Rev. {\bf D80}, 015023 (2009), 0903.5151.

\bibitem{Davidson:2010xv}
S.~Davidson and G.~J. Grenier,
\newblock Phys.Rev. {\bf D81}, 095016 (2010), 1001.0434.

\bibitem{Goudelis:2011un}
A.~Goudelis, O.~Lebedev, and J.-h. Park,
\newblock Phys.Lett. {\bf B707}, 369 (2012), 1111.1715.

\bibitem{Harnik:2012pb}
R.~Harnik, J.~Kopp, and J.~Zupan,
\newblock JHEP {\bf 1303}, 026 (2013), 1209.1397.

\bibitem{Davidson:2012ds}
S.~Davidson and P.~Verdier,
\newblock Phys.Rev. {\bf D86}, 111701 (2012), 1211.1248.

\bibitem{Crivellin:2013wna}
A.~Crivellin, A.~Kokulu, and C.~Greub,
\newblock Phys.Rev. {\bf D87}, 094031 (2013), 1303.5877.

\bibitem{Dery:2013rta}
A.~Dery, A.~Efrati, Y.~Hochberg, and Y.~Nir,
\newblock JHEP {\bf 1305}, 039 (2013), 1302.3229.

\bibitem{Arhrib:2012ax}
A.~Arhrib, Y.~Cheng, and O.~C. Kong,
\newblock Phys.Rev. {\bf D87}, 015025 (2013), 1210.8241.

\bibitem{Arana-Catania:2013xma}
M.~Arana-Catania, E.~Arganda, and M.~Herrero,
\newblock (2013), 1304.3371.

\bibitem{Buchmuller:1985jz}
W.~Buchmuller and D.~Wyler,
\newblock Nucl.Phys. {\bf B268}, 621 (1986).

\bibitem{Grzadkowski:2010es}
B.~Grzadkowski, M.~Iskrzynski, M.~Misiak, and J.~Rosiek,
\newblock JHEP {\bf 1010}, 085 (2010), 1008.4884.

\bibitem{Gunion:1989we}
J.~F. Gunion, H.~E. Haber, G.~L. Kane, and S.~Dawson,
\newblock Front.Phys. {\bf 80}, 1 (2000).

\bibitem{Branco:2011iw}
G.~Branco {\em et~al.},
\newblock Phys.Rept. {\bf 516}, 1 (2012), 1106.0034.

\bibitem{Chang:1993kw}
D.~Chang, W.~Hou, and W.-Y. Keung,
\newblock Phys.Rev. {\bf D48}, 217 (1993), hep-ph/9302267.

\bibitem{Petrov:2013vka}
A.~A. Petrov and D.~V. Zhuridov,
\newblock (2013), 1308.6561.

\bibitem{Daub:2012mu}
J.~Daub, H.~Dreiner, C.~Hanhart, B.~Kubis, and U.~Meissner,
\newblock JHEP {\bf 1301}, 179 (2013), 1212.4408.

\bibitem{Amendolia:1986wj}
NA7 Collaboration, S.~Amendolia {\em et~al.},
\newblock Nucl.Phys. {\bf B277}, 168 (1986).

\bibitem{Aloisio:2004bu}
KLOE Collaboration, A.~Aloisio {\em et~al.},
\newblock Phys.Lett. {\bf B606}, 12 (2005), hep-ex/0407048.

\bibitem{Akhmetshin:2006bx}
CMD-2 Collaboration, R.~Akhmetshin {\em et~al.},
\newblock Phys.Lett. {\bf B648}, 28 (2007), hep-ex/0610021.

\bibitem{Aubert:2009ad}
BaBar Collaboration, B.~Aubert {\em et~al.},
\newblock Phys.Rev.Lett. {\bf 103}, 231801 (2009), 0908.3589.

\bibitem{Ambrosino:2010bv}
KLOE Collaboration, F.~Ambrosino {\em et~al.},
\newblock Phys.Lett. {\bf B700}, 102 (2011), 1006.5313.

\bibitem{Anderson:1999ui}
CLEO Collaboration, S.~Anderson {\em et~al.},
\newblock Phys.Rev. {\bf D61}, 112002 (2000), hep-ex/9910046.

\bibitem{Fujikawa:2008ma}
Belle Collaboration, M.~Fujikawa {\em et~al.},
\newblock Phys.Rev. {\bf D78}, 072006 (2008), 0805.3773.

\bibitem{Barate:1997hv}
ALEPH Collaboration, R.~Barate {\em et~al.},
\newblock Z.Phys. {\bf C76}, 15 (1997).

\bibitem{Gounaris:1968mw}
G.~Gounaris and J.~Sakurai,
\newblock Phys.Rev.Lett. {\bf 21}, 244 (1968).

\bibitem{Gasser:1983yg}
J.~Gasser and H.~Leutwyler,
\newblock Annals Phys. {\bf 158}, 142 (1984).

\bibitem{Truong:1988zp}
T.~N. Truong,
\newblock Phys.Rev.Lett. {\bf 61}, 2526 (1988).

\bibitem{Kuhn:1990ad}
J.~H. Kuhn and A.~Santamaria,
\newblock Z.Phys. {\bf C48}, 445 (1990).

\bibitem{Gasser:1990bv}
J.~Gasser and U.~G. Meissner,
\newblock Nucl.Phys. {\bf B357}, 90 (1991).

\bibitem{Colangelo:1996hs}
G.~Colangelo, M.~Finkemeier, and R.~Urech,
\newblock Phys.Rev. {\bf D54}, 4403 (1996), hep-ph/9604279.

\bibitem{Dung:1996rp}
L.~v. Dung and T.~N. Truong,
\newblock (1996), hep-ph/9607378.

\bibitem{Hannah:1996tp}
T.~Hannah,
\newblock Phys.Rev. {\bf D54}, 4648 (1996), hep-ph/9611307.

\bibitem{Hannah:1997ux}
T.~Hannah,
\newblock Phys.Rev. {\bf D55}, 5613 (1997), hep-ph/9701389.

\bibitem{Guerrero:1997ku}
F.~Guerrero and A.~Pich,
\newblock Phys.Lett. {\bf B412}, 382 (1997), hep-ph/9707347.

\bibitem{Bijnens:1998fm}
J.~Bijnens, G.~Colangelo, and P.~Talavera,
\newblock JHEP {\bf 9805}, 014 (1998), hep-ph/9805389.

\bibitem{Oller:2000ug}
J.~Oller, E.~Oset, and J.~Palomar,
\newblock Phys.Rev. {\bf D63}, 114009 (2001), hep-ph/0011096.

\bibitem{DeTroconiz:2001wt}
J.~De~Troconiz and F.~Yndurain,
\newblock Phys.Rev. {\bf D65}, 093001 (2002), hep-ph/0106025.

\bibitem{Pich:2001pj}
A.~Pich and J.~Portoles,
\newblock Phys.Rev. {\bf D63}, 093005 (2001), hep-ph/0101194.

\bibitem{Ananthanarayan:2011xt}
B.~Ananthanarayan, I.~Caprini, and I.~S. Imsong,
\newblock Phys.Rev. {\bf D83}, 096002 (2011), 1102.3299.

\bibitem{Hanhart:2012wi}
C.~Hanhart,
\newblock Phys.Lett. {\bf B715}, 170 (2012), 1203.6839.

\bibitem{Jamin:2006tk}
M.~Jamin, A.~Pich, and J.~Portoles,
\newblock Phys.Lett. {\bf B640}, 176 (2006), hep-ph/0605096.

\bibitem{Jamin:2008qg}
M.~Jamin, A.~Pich, and J.~Portoles,
\newblock Phys.Lett. {\bf B664}, 78 (2008), 0803.1786.

\bibitem{Boito:2010me}
D.~Boito, R.~Escribano, and M.~Jamin,
\newblock JHEP {\bf 1009}, 031 (2010), 1007.1858.

\bibitem{Bernard:2011ae}
V.~Bernard, D.~Boito, and E.~Passemar,
\newblock Nucl.Phys.Proc.Suppl. {\bf 218}, 140 (2011), 1103.4855.

\bibitem{Dumm:2013zh}
D.~G. Dumm and P.~Roig,
\newblock (2013), 1301.6973.

\bibitem{Descotes-Genon:2013uya}
S.~Descotes-Genon, E.~Kou, and B.~Moussallam,
\newblock (2013), 1303.2879.

\bibitem{Escribano:2013bca}
R.~Escribano, S.~González-Solís, and P.~Roig,
\newblock (2013), 1307.7908.

\bibitem{BBP}
V.~Bernard, D.~Boito, and E.~Passemar, 2013,
\newblock Work in preparation.

\bibitem{Watson:1952ji}
K.~M. Watson,
\newblock Phys.Rev. {\bf 88}, 1163 (1952).

\bibitem{Ananthanarayan:2000ht}
B.~Ananthanarayan, G.~Colangelo, J.~Gasser, and H.~Leutwyler,
\newblock Phys.Rept. {\bf 353}, 207 (2001), hep-ph/0005297.

\bibitem{CCL}
I.~Caprini, G.~Colangelo, and H.~Leutwyler, 2013,
\newblock Work in preparation.

\bibitem{Ecker:1988te}
G.~Ecker, J.~Gasser, A.~Pich, and E.~de~Rafael,
\newblock Nucl.Phys. {\bf B321}, 311 (1989).

\bibitem{Ecker:1989yg}
G.~Ecker, J.~Gasser, H.~Leutwyler, A.~Pich, and E.~de~Rafael,
\newblock Phys.Lett. {\bf B223}, 425 (1989).

\bibitem{Cirigliano:2004ue}
V.~Cirigliano, G.~Ecker, M.~Eidemuller, A.~Pich, and J.~Portoles,
\newblock Phys.Lett. {\bf B596}, 96 (2004), hep-ph/0404004.

\bibitem{Bernard:2009zm}
V.~Bernard, M.~Oertel, E.~Passemar, and J.~Stern,
\newblock Phys.Rev. {\bf D80}, 034034 (2009), 0903.1654.

\bibitem{Lepage:1980fj}
G.~P. Lepage and S.~J. Brodsky,
\newblock Phys.Rev. {\bf D22}, 2157 (1980).

\bibitem{Donoghue:1990xh}
J.~F. Donoghue, J.~Gasser, and H.~Leutwyler,
\newblock Nucl.Phys. {\bf B343}, 341 (1990).

\bibitem{Moussallam:1999aq}
B.~Moussallam,
\newblock Eur.Phys.J. {\bf C14}, 111 (2000), hep-ph/9909292.

\bibitem{Au:1986vs}
K.~Au, D.~Morgan, and M.~Pennington,
\newblock Phys.Rev. {\bf D35}, 1633 (1987).

\bibitem{Muskhelishvili}
N.~Muskhelishvili, 1953,
\newblock Singular integral equations.

\bibitem{Omnes:1958hv}
R.~Omnes,
\newblock Nuovo Cim. {\bf 8}, 316 (1958).

\bibitem{Roy:1971tc}
S.~Roy,
\newblock Phys.Lett. {\bf B36}, 353 (1971).

\bibitem{GarciaMartin:2011cn}
R.~Garcia-Martin, R.~Kaminski, J.~Pelaez, J.~Ruiz~de Elvira, and F.~Yndurain,
\newblock Phys.Rev. {\bf D83}, 074004 (2011), 1102.2183.

\bibitem{Buettiker:2003pp}
P.~Buettiker, S.~Descotes-Genon, and B.~Moussallam,
\newblock Eur.Phys.J. {\bf C33}, 409 (2004), hep-ph/0310283.

\bibitem{Moussallam:2011zg}
B.~Moussallam,
\newblock Eur.Phys.J. {\bf C71}, 1814 (2011), 1110.6074.

\bibitem{Colangelo:2010et}
G.~Colangelo {\em et~al.},
\newblock Eur.Phys.J. {\bf C71}, 1695 (2011), 1011.4408.

\bibitem{Bernard:2012fw}
V.~Bernard, S.~Descotes-Genon, and G.~Toucas,
\newblock JHEP {\bf 1206}, 051 (2012), 1203.0508.

\bibitem{Shifman:1978zn}
M.~A. Shifman, A.~Vainshtein, and V.~I. Zakharov,
\newblock Phys.Lett. {\bf B78}, 443 (1978).

\bibitem{Cheng:1987rs}
T.~Cheng and M.~Sher,
\newblock Phys.Rev. {\bf D35}, 3484 (1987).

\bibitem{Celis:2013rcs}
A.~Celis, V.~Ilisie, and A.~Pich,
\newblock JHEP {\bf 1307}, 053 (2013), 1302.4022.

\bibitem{Miyazaki:2011xe}
Belle Collaboration, Y.~Miyazaki {\em et~al.},
\newblock Phys.Lett. {\bf B699}, 251 (2011), 1101.0755.

\bibitem{Miyazaki:2012mx}
Belle Collaboration, Y.~Miyazaki {\em et~al.},
\newblock Phys.Lett. {\bf B719}, 346 (2013), 1206.5595.

\bibitem{Bonvicini:1997bw}
CLEO Collaboration, G.~Bonvicini {\em et~al.},
\newblock Phys.Rev.Lett. {\bf 79}, 1221 (1997), hep-ex/9704010.

\bibitem{Aubert:2009ag}
BaBar Collaboration, B.~Aubert {\em et~al.},
\newblock Phys.Rev.Lett. {\bf 104}, 021802 (2010), 0908.2381.

\bibitem{Hayasaka:2010np}
K.~Hayasaka {\em et~al.},
\newblock Phys.Lett. {\bf B687}, 139 (2010), 1001.3221.

\bibitem{Altmannshofer:2012ar}
W.~Altmannshofer, S.~Gori, and G.~D. Kribs,
\newblock Phys.Rev. {\bf D86}, 115009 (2012), 1210.2465.

\bibitem{Barger:2013ofa}
V.~Barger, L.~L. Everett, H.~E. Logan, and G.~Shaughnessy,
\newblock (2013), 1308.0052.

\bibitem{Lopez-Val:2013yba}
D.~Lopez-Val, T.~Plehn, and M.~Rauch,
\newblock (2013), 1308.1979.

\bibitem{Feldmann:1998vh}
T.~Feldmann, P.~Kroll, and B.~Stech,
\newblock Phys.Rev. {\bf D58}, 114006 (1998), hep-ph/9802409.

\bibitem{Beneke:2002jn}
M.~Beneke and M.~Neubert,
\newblock Nucl.Phys. {\bf B651}, 225 (2003), hep-ph/0210085.

\bibitem{Gross:1979ur}
D.~J. Gross, S.~Treiman, and F.~Wilczek,
\newblock Phys.Rev. {\bf D19}, 2188 (1979).

\bibitem{Aubert:2006cz}
BaBar Collaboration, B.~Aubert {\em et~al.},
\newblock Phys.Rev.Lett. {\bf 98}, 061803 (2007), hep-ex/0610067.

\bibitem{Aad:2012yfa}
ATLAS Collaboration, G.~Aad {\em et~al.},
\newblock JHEP {\bf 1302}, 095 (2013), 1211.6956.

\bibitem{Harlander:2012pb}
R.~V. Harlander, S.~Liebler, and H.~Mantler,
\newblock Computer Physics Communications {\bf 184}, 1605 (2013), 1212.3249.

\bibitem{Harlander:2002wh}
R.~V. Harlander and W.~B. Kilgore,
\newblock Phys.Rev.Lett. {\bf 88}, 201801 (2002), hep-ph/0201206.

\bibitem{Eriksson:2009ws}
D.~Eriksson, J.~Rathsman, and O.~Stal,
\newblock Comput.Phys.Commun. {\bf 181}, 189 (2010), 0902.0851.

\bibitem{Assamagan:2002kf}
K.~A. Assamagan, A.~Deandrea, and P.-A. Delsart,
\newblock Phys.Rev. {\bf D67}, 035001 (2003), hep-ph/0207302.

\end{thebibliography}

\end{document}